\documentclass[aps,prb,twocolumn,10pt,floatfix]{revtex4-1}

\usepackage[pdftex]{graphicx} 
\usepackage{epstopdf}
\usepackage{verbatim}
\usepackage{subfigure}
\usepackage{tabularx}
\usepackage{amsmath,amssymb,amsfonts}
\usepackage{wasysym}
\usepackage{bm}
\usepackage{bbm}
\usepackage{dsfont}
\usepackage{float}
\usepackage{nicefrac}
\usepackage{scrextend}
\usepackage{mathtools}

\usepackage[usenames]{color}
\usepackage[bookmarks=true,colorlinks,linkcolor=blue, urlcolor=blue,citecolor=blue]{hyperref}

\definecolor{darkred}{rgb}{0.847059, 0.141176, 0.164706}
\definecolor{darkgreen}{rgb}{0,0.4,0}
\definecolor{darkblue}{rgb}{0.254902, 0.411765, 0.882353}

\newcommand{\eqnref}[1]{Eq.~(\ref{#1})}
\newcommand{\nn}{\nonumber\\}

\newcommand{\ket}[1]{\ensuremath{|#1\rangle}}

\newcommand{\vev}[1]{\left\langle#1\right\rangle}
\newcommand{\mc}{\mathcal}
\newcommand{\mb}{\mathbb}
\newcommand{\bs}{\boldsymbol}
\newcommand{\rx}{\right)}
\newcommand{\lx}{\left(}
\newcommand{\rz}{\right]}
\newcommand{\lz}{\left[}

\newcommand{\dg}{\dagger}
\newcommand{\pg}{{\phantom{\dagger}}}

\newcommand{\ua}{\uparrow}
\newcommand{\da}{\downarrow}
\newcommand{\p}{\prime}

\newcolumntype{C}[1]{>{\centering\let\newline\\\arraybackslash\hspace{0pt}}m{#1}}

\begin{document}

\title{Non-Kitaev spin liquids in Kitaev materials}

\author{Yao-Dong Li$^{1,2}$}
\author{Xu Yang$^{3,5}$}
\author{Yi Zhou$^4$}
\author{Gang Chen$^{1,5}$}
\thanks{Currently on leave from Fudan University, China}
\email{gangchen.physics@gmail.com}
\affiliation{$^{1}$State Key Laboratory of Surface Physics and Department of Physics, 
Fudan University, Shanghai 200433, China}
\affiliation{$^{2}$Department of Physics, 
University of California Santa Barbara, CA 93106, United States}
\affiliation{$^{3}$Department of Physics, Boston College, Chestnut Hill, Massachusetts 02467, United States}
\affiliation{$^{4}$Institute of Physics, Chinese Academy of Sciences, Beijing 100190, China}
\affiliation{$^{5}$Department of Physics and Center of Theoretical and Computational Physics,
The University of Hong Kong, Pokfulam Road, Hong Kong, China}
\date{\today}

\begin{abstract}
We point out that the Kitaev materials may not necessarily support Kitaev spin liquid.
It is well-known that having a Kitaev term in the spin interaction is not
the sufficient condition for the Kitaev spin liquid ground state.
Many other spin liquids may be stabilized by the competing
spin interactions of the systems. We thus explore the possibilities
of non-Kitaev spin liquids in the honeycomb Kitaev materials.
We carry out a systematic classification of gapped $\mb{Z}_2$ spin liquids
using the Schwinger boson representation for the spin variables.
The presence of strong spin-orbit coupling in the Kitaev materials
brings new ingredients into the projective symmetry group classification
of the non-Kitaev spin liquid. We predict the spectroscopic properties
of these gapped non-Kitaev spin liquids. Moreover, among the
gapped spin liquids that we discover, we identify the spin liquid
whose spinon condensation leads to the zig-zag magnetic order that
was observed in Na$_2$IrO$_3$ and $\alpha$-RuCl$_3$.
We further discuss the possibility of gapped $\mathbb{Z}_2$ spin liquid
and the deconfined quantum criticality from the zig-zag magnetic order to
spin dimerization
in pressurized $\alpha$-RuCl$_3$.
\end{abstract}

\maketitle


\section{Introduction}
\label{sec1}

Kitaev spin liquid was proposed by A. Kitaev
when he constructed an elegant model and solved it exactly~\cite{KitaevHeisenberg}.
An interesting connection to Na$_2$IrO$_3$ was made
by G. Jackeli and G. Khaliulin~\cite{Jackeli2009}. 
It was shown that the strong spin-orbit
coupling (SOC) of iridium electrons could give rise to a Kitaev interaction
in the effective spin Hamiltonian for the ${j=1/2}$ iridium local 
moments. Since then, many iridates were synthesized and 
explored~\cite{Takagi,PhysRevB.91.235147,PhysRevLett.109.266406,PhysRevB.88.085125,PhysRevB.85.180403,PhysRevB.87.220407,PhysRevLett.110.076402,PhysRevB.83.220403,PhysRevLett.114.077202,Modic}, 
including the recent $\alpha$-RuCl$_3$~\cite{Binotto,PhysRevB.53.12769,PhysRevB.90.041112,PhysRevB.93.075144,RuCl3,Banerjee1055} 
and the very early hyperkagome lattice spin liquid material
Na$_4$Ir$_3$O$_8$~\cite{Na4Ir3O8} where the ${j=1/2}$ local moment~\cite{KimBJ}
and the anisotropic spin interaction were proposed~\cite{Chen2008}. 
These materials are dubbed ``Kitaev materials'' and have sparked an
active search of Kitaev spin liquid~\cite{Trebst,RevModPhys.89.025003,Moessner,
doi:10.1146/annurev-conmatphys-033117-053934,BalentsS}.

Generally speaking, the list of Kitaev materials goes beyond
iridates and ruthenates~\cite{PhysRevB.95.085132,PhysRevB.98.054408,Jang}.
What gives the Kitaev interaction is the strong SOC, and
this is common to magnetic materials with heavy atoms.
Therefore, any strong spin-orbit-coupled Mott insulator with
spin-orbit-entangled effective spin-1/2 moments and a proper
lattice geometry can be a Kitaev material.
In particular, the rare-earth magnets, that have the same lattice
structure as iridates and ruthenates, could be ideal Kitaev 
materials~\cite{PhysRevB.95.085132}. Despite the growing list of Kitaev materials,
all these systems face one crucial issue---there are many competing
interactions that coexist with the Kitaev interaction.
For example, for the nearest-neighbor bonds in Na$_2$IrO$_3$ and $\alpha$-RuCl$_3$,
three extra interactions beyond the Kitaev interaction are present~\cite{PhysRevLett.112.077204}, 
not to mention many further neighbor (anisotropic) spin interactions that arise
from the large spatial extension of the $4d$/$5d$ electron wavefunction.

In fact it has been shown that Kitaev spin liquid is fragile and small
perturbation could actually destabilze it~\cite{PhysRevB.83.245104,
PhysRevLett.110.097204,PhysRevLett.118.137203,PhysRevX.5.041035,PhysRevB.90.155126}. 
Meanwhile, the real materials
contain many competing interactions that may be as important as the
Kitaev interaction, the candidate quantum spin liquids (QSLs) for 
these materials remain to be examined. On the other hand,
for any other gapped QSL that is not Kitaev spin liquid,
if it is realized, it will be stable against
small local perturbations regardless of the Kitaev interaction.
This means that having the Kitaev interaction in the Hamiltonian
is insufficient to induce Kitaev spin liquid and other
competing interactions could instead favor different QSL 
ground states. For example, the $J_1$-$J_2$ spin-1/2 Heisenberg
model on the honeycomb lattice in certain parameter regime was
proposed to support a gapped QSL that is clearly not a Kitaev 
spin liquid~\cite{PhysRevB.88.165138}.

In this work, we deviate from the ``hot spot'' of searching for
Kitaev spin liquid in Kitaev materials. Instead, our goal here is
to find possible candidate QSLs in Kitaev materials that are not
Kitaev spin liquid and to predict the experimental consequences of 
them. Considering the richness of Kitaev materials, it is very likely 
that these non-Kitaev QSLs may actually be stabilized in 
certain systems. A recent study of pressurized $\alpha$-RuCl$_3$
indeed suggested some evidence for a possible QSL~\cite{PhysRevB.97.245149}.
This experimental work motivates us to search for non-Kitaev QSLs in 
these systems. We carry out a systematic projective symmetry group (PSG) 
classification of gapped $\mb{Z}_2$ QSLs on a honeycomb lattice using 
Schwinger boson~\cite{Read1991,WenPSGPLA, WenPSGPRB,WangAshvin,FaWang}
representation of the spins. Due to the spin-orbit-entangled nature
of the local moments, the symmetry transformation operates both on
the spin components and on the spin 
position~\cite{PhysRevB.96.054445,PhysRevB.90.174417,PhysRevB.88.174405}. 
This new symmetry property
gives a different classification scheme from the existing PSG analysis.
From the PSG results, we predict the spectroscopic properties of
different $\mb{Z}_2$ QSLs on the honeycomb lattice.
Moreover, we study the proximate magnetic orders out of
the QSLs by condensing the spinons~\cite{Read1991,Sachdev1992}. 
The magnetic wavevector of the zig-zag magnetic order, that was observed in Na$_2$IrO$_3$ 
and $\alpha$-RuCl$_3$~\cite{PhysRevB.85.180403,PhysRevLett.110.097204,PhysRevB.91.144420}, 
naturally connects with the $\mathbb{Z}$2B QSLs via the spinon condensation.

The remaining parts of the paper are organized as follows.
In Sec.~\ref{sec2}, we introduce the Schwinger boson construction
for the $\mathbb{Z}_2$ QSLs with spin-orbit-entangled local moments.
In Sec.~\ref{sec3}, we explain the specific properties of the
symmetry operations under the Schwinger boson framework.
In Sec.~\ref{sec4}, we obtain 16 distinct $\mathbb{Z}_2$ QSLs 
from the PSG classifications and study the phase diagram of 
several representative mean-field QSL states.
In Sec.~\ref{sec5}, we explore the spectroscopic properties and
the proximate magnetic phases of the aforementioned mean-field 
$\mathbb{Z}_2$ QSLs. Finally in Sec.~\ref{sec6},
we discuss the relevant experiments and especially explain the 
possibilities for the pressurized $\alpha$-RuCl$_3$.

\section{Schwinger boson construction}
\label{sec2}

The gapped $\mb{Z}_2$ spin liquids can be studied by either Schwinger
boson or Abrikosov fermion approach. We here adopt the Schwinger boson 
construction since it is easier to explore the proximate magnetic orders 
with bosonic variables. In the Schwinger boson representation,
the effective spin $\bs{S}_{i}$ on site $i$ is given by ${\bs{S}_{i}^{} =
\frac{1}{2} b_{i\alpha}^\dg \bs{\sigma}_{\alpha\beta}^\pg b_{i\beta}^\pg}$
where $b_{i\alpha}$ (${\alpha = \uparrow, \downarrow}$) is the bosonic
spinon operator. The Hilbert space is enlarged due to the introduction
of the spinons; to project out unphysical states, the constraint
${\sum_\alpha b^\dg_{i\alpha} b^\pg_{i\alpha} = 1}$ on local
boson number is imposed. The most general candidate
mean-field Hamiltonian for the $\mb{Z}_2$ spin liquids has
the following form,
\begin{eqnarray} 
\label{Hmf}
H_{\text{MF}} &=& \sum_{\vev{ij},\alpha\beta}
(     u^{\text A}_{ij,\alpha\beta} b^\dg_{i\alpha} b^\pg_{j\beta}
    + u^{\text B}_{ij,\alpha\beta} b^\pg_{i\alpha} b^\pg_{j\beta}
    + h.c.
) \nn
&&
+ \sum_{i} \mu_{i}^{} (\sum_{\alpha} b^\dg_{i \alpha} b^\pg_{i \alpha} - 1)
\end{eqnarray}
where we have restricted the mean-field ansatz to nearest
neighbors and introduced the chemical potential $\mu_{i}$
to enforce the boson number constraint and we have used the 
superscript A/B to represent hopping/pairing terms in the coefficients $u$.
Due to the spin-orbit-entangled nature of the local moments,
the SU(2) symmetry breaking terms exist in the mean-field ansatz.
Using the hermiticity of the Hamiltonian and bosonic statistics of the spinons, 
it is easy to show that
${u^B_{ij,\ua\da} = u^B_{ji,\da\ua}}$,
${u^B_{ij,\alpha\alpha} = u^B_{j i,\alpha\alpha}}$,
${u^A_{ij,\alpha\alpha} = (u^A_{j i,\alpha\alpha})^\ast}$,
and ${u^A_{ij,\ua\da} = (u^A_{j i,\da\ua})^\ast}$.

\begin{figure}[b]
\centering
\includegraphics[width=5cm]{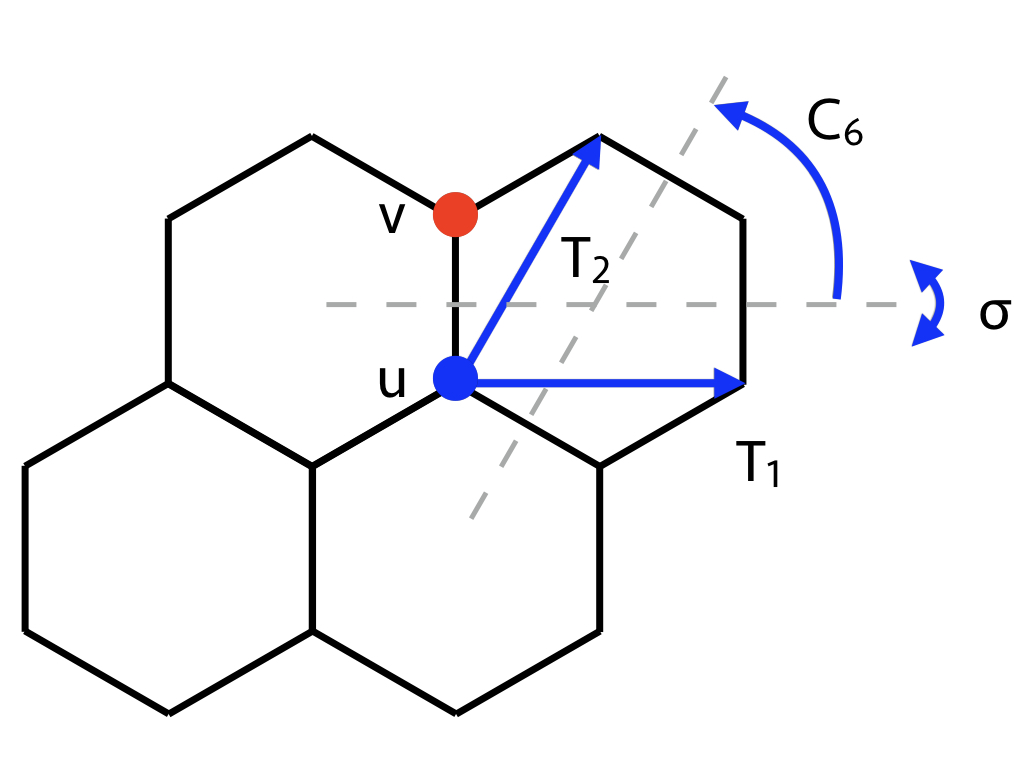}
\caption{The honeycomb lattice and its symmetries. 
Blue/red circles indicate the two sublattices denoted as $u/v$. 
The space-group generators are translations $T_1$ and $T_2$, 
sixfold rotation $C_6$ around the plaquette center, and 
horizontal reflection $\sigma$ through the hexagon center.
}
\label{fig1}
\end{figure}

\section{Projective symmetry group}
\label{sec3}

In this section we follow the projective symmetry group (PSG) approach introduced 
in Refs.~\onlinecite{WenPSGPLA, WenPSGPRB} to classify the spinon mean field states 
based on the symmetries of the honeycomb layers of Kitaev materials. The 
spinon mean field state will be a reasonable description of the QSLs, 
provided the QSL survives the quantum fluctuations beyond mean field.

The physical symmetry group of the Hamiltonian contains both space group symmetries and the time-reversal symmetry.
For simplicity, we fix the representation of the time-reversal symmetry to be the following throughout the paper:
\begin{eqnarray} 
\mc{T} : b_{i\ua} \to b_{i\da},\quad b_{i\da} \to -b_{i\da}.
\end{eqnarray}
The space group symmetries, on the other hand, can be represented projectively by the spinons.
Therefore, we will only take the space group symmetries into account for the PSG classification; the time-reversal symmetry commutes with all the space group symmetries and does not affect the classification (see Appendix~\ref{appB}).
The time-reversal symmetry will nevertheless restrict the form of the mean-field Hamiltonian (see Appendix~\ref{appC}).

The lattice system of the honeycomb layer is shown in Fig.~\ref{fig1} 
and defined in Appendix~\ref{appA}. The space group is generated by two 
translations $T_1$ and $T_2$, a counterclockwise sixfold rotation $C_6$ 
around the hexagon center, and a reflection $\sigma$ around the horizontal 
axis through the same hexagon center. Under the symmetry operation $\mc{O}$, 
the bosonic spinon transforms as
\begin{eqnarray} 
\label{psg}
    b^\pg_{i} \to \hat{\mc{O}}^\dg b_i \hat{\mc{O}} 
    = \mc{G}^\mc{O}_{\mc{O}(i)} \, \mc{U}^\pg_\mc{O} \, b^\pg_{\mc{O}(i)}
\end{eqnarray}
where ${\mc{G}^\mc{O}_{\mc{O}(i)} = e^{i \phi_{\mc{O}}[\mc{O}(i)]}}$ is 
a local U(1) gauge transformation, which leaves the spin operators invariant. 
The gauge transformation is generally nontrivial, hence incorporated in the 
symmetry operation in Eq.~\eqref{psg}. After projection into the physical Hilbert space, 
spinons states related by a pure gauge transformation should give the same physical state.  
In Eq.~\eqref{psg} we have introduced the spin rotation $\mc{U}_\mc{O}$ to account for 
the effects of SOC, which rotates the position and spin simultaneously. In explicit forms, 
we have 
${\mc{U}_{T_1} = \mc{U}_{T_2} = {\bf 1}_{2 \times 2}},
 {\mc{U}_{C_6} = \exp \lx i \frac{\pi}{3} \frac{\sigma_z}{2} \rx} ,
 {\mc{U}_{\sigma} = \exp \lx i \pi \frac{\sigma_x}{2} \rx}$.

For mean-field Hamiltonian of the form in \eqnref{Hmf} to be invariant 
under the symmetry transformation $\mc{O}$, the coefficients should satisfy
\begin{eqnarray}
    u^A_{\mc{O}(i) \mc{O}(j), \alpha\beta} &=&
    \lx \mc{G}^\mc{O}_{\mc{O}(i)} \rx^\ast
    \mc{G}^\mc{O}_{\mc{O}(j)}
    \lx \mc{U}^\ast_\mc{O} \rx_{\alpha\nu}
    \lx \mc{U}_\mc{O} \rx_{\beta\lambda} u^A_{i j, \nu\lambda}, \\
    u^B_{\mc{O}(i) \mc{O}(j), \alpha\beta} &=&
    \mc{G}^\mc{O}_{\mc{O}(i)}
    \mc{G}^\mc{O}_{\mc{O}(j)}
    \lx \mc{U}_\mc{O} \rx_{\alpha\nu}
    \lx \mc{U}_\mc{O} \rx_{\beta\lambda} u^B_{i j, \nu\lambda},
\end{eqnarray}
where we have used the fact that $\mc{U}_\mc{O}$ commutes with $\mc{G}^\mc{O}$.
For a general pair of sites $(i, j)$, the above equations are solvable if for 
each group relation ${\mc{O}_1 \mc{O}_2 \cdots \mc{O}_n = 1}$, the following 
identities are satisfied,
\begin{eqnarray} 
\label{psg_eqn}
    &{\mc{U}_{\mc{O}_1} \mc{U}_{\mc{O}_2}   \cdots \mc{U}_{\mc{O}_n} \mc{G}^{\mc{O}_1}_i
    \mc{G}^{\mc{O}_2}_{\mc{O}_2\mc{O}_3\cdots \mc{O}_n (i)}
    \mc{G}^{\mc{O}_3}_{\mc{O}_3\cdots \mc{O}_n (i)}   \cdots 
    \mc{G}^{\mc{O}_n}_{\mc{O}_n (i)} = \pm 1} \nn
    &\Leftrightarrow
    \mc{G}^{\mc{O}_1}_i
    \mc{G}^{\mc{O}_2}_{\mc{O}_2\mc{O}_3\cdots \mc{O}_n (i)}
    \mc{G}^{\mc{O}_3}_{\mc{O}_3\cdots \mc{O}_n (i)}  \cdots 
    \mc{G}^{\mc{O}_n}_{\mc{O}_n (i)} = \pm 1,
\end{eqnarray}
where $\pm 1$ is either element of $\mb{Z}_2$, the invariant gauge group (IGG). 
The IGG turns out to be the gauge group of the low-energy effective theory of 
the QSL state \cite{WenPSGPLA, WenPSGPRB}. Here, since we are considering 
$\mb{Z}_2$ QSLs, the IGG should also be $\mb{Z}_2$. The two lines in 
Eq.~\eqref{psg_eqn} are equivalent because the identity element involves either 
rotation by $0$ or $2\pi$, 
so $\mc{U}_{\mc{O}_1} \mc{U}_{\mc{O}_2} \cdots \mc{U}_{\mc{O}_n}= \pm 1$, 
and the group relations constraint only the phases $\phi_\mc{O}$. Given the 
defining relations between group generators $T_1,T_2,C_6,\sigma$, we can solve 
for all the possible gauge transformation functions $\phi_{\mc{O}}(i)$'s 
compatible with Eq.~\eqref{psg_eqn}.

\begin{table}
\renewcommand{\arraystretch}{1.2}
	\begin{tabular}{c|c c c c c}
	\hline\hline
$\mathbb{Z}_2$ QSL & $\mc{G}^{T_1}$ & $\mc{G}^{T_2}$ & $\mc{G}^{C_6}$ & $\mc{G}^\sigma[u]$ & $\mc{G}^\sigma[v]$\\\hline
		$\mathbb{Z}$2A000 &  $1$         & $1$          & $1$          & $1$       & $1$ \\
		$\mathbb{Z}$2A001 &  $1$         & $1$          & $i$          & $i$       & $-i$\\
		$\mathbb{Z}$2A010 &  $1$         & $1$          & $i$          & $1$       & $1$ \\
		$\mathbb{Z}$2A011 &  $1$         & $1$          & $-1$         & $i$       & $-i$\\
		$\mathbb{Z}$2A100 &  $1$         & $1$          & $i$          & $i$       & $i$\\
		$\mathbb{Z}$2A101 &  $1$         & $1$          & $-1$         & $-1$      & $1$ \\
		$\mathbb{Z}$2A110 &  $1$         & $1$          & $-1$         & $i$       & $i$\\
		$\mathbb{Z}$2A111 &  $1$         & $1$          & $-i$         & $-1$      & $1$\\
		$\mathbb{Z}$2B000 &  $1$         & $(-1)^x$     & $i^{x(x+2y-1)}$& $i^{2x+y(y+1)}$       & $i^{2x+y(y+1)}$ \\
		$\mathbb{Z}$2B001 &  $1$         & $(-1)^x$     & $i^{x(x+2y-1)+1}$& $i^{2x+y(y+1)+1}$       & $i^{2x+y(y+1)-1}$ \\
		$\mathbb{Z}$2B010 &  $1$         & $(-1)^x$     & $i^{x(x+2y-1)+1}$& $i^{2x+y(y+1)}$       & $i^{2x+y(y+1)}$ \\
		$\mathbb{Z}$2B011 &  $1$         & $(-1)^x$     & $i^{x(x+2y-1)+2}$& $i^{2x+y(y+1)+1}$       & $i^{2x+y(y+1)-1}$ \\
		$\mathbb{Z}$2B100 &  $1$         & $(-1)^x$     & $i^{x(x+2y-1)+1}$& $i^{2x+y(y+1)+1}$       & $i^{2x+y(y+1)+1}$ \\
		$\mathbb{Z}$2B101 &  $1$         & $(-1)^x$     & $i^{x(x+2y-1)+2}$& $i^{2x+y(y+1)+2}$       & $i^{2x+y(y+1)}$ \\
		$\mathbb{Z}$2B110 &  $1$         & $(-1)^x$     & $i^{x(x+2y-1)+2}$& $i^{2x+y(y+1)+1}$       & $i^{2x+y(y+1)+1}$ \\
		$\mathbb{Z}$2B111 &  $1$         & $(-1)^x$     & $i^{x(x+2y-1)+3}$& $i^{2x+y(y+1)+2}$       & $i^{2x+y(y+1)}$ \\
	\hline\hline
	\end{tabular}
	\caption{List of the gauge transformations associated with the
		symmetry operations of the 16 $\mb{Z}_2$ QSLs,
		where $(x,y,w)$ denotes the
		site in the honeycomb coordinate system.}
	\label{stab1}
\end{table}

\section{The 16 classes of $\mb{Z}_2$ QSLs and the mean-field phase diagram}
\label{sec4}

The solutions of the $\phi_\mc{O}$'s for equations of the form in \eqnref{psg_eqn} are as follows:
\begin{eqnarray}
&& \phi_{T_1}(x,y,w) = 0,   \\
&& \phi_{T_2}(x,y,w) = p_1 \pi x,   \\
&& \phi_{C_6}(x,y,w) = \frac{\pi}{2} \lx p_1 x (x+2y-1) + p_7 + p_8 + p_9\rx,   \\
&& \phi_{\sigma}(x,y,u) = \frac{\pi}{2} \lx 2p_1x+p_1y(y+1)+p_7+p_9 \rx,   \\
&& \phi_{\sigma}(x,y,v) = \frac{\pi}{2} \lx 2p_1x+p_1y(y+1)+p_7-p_9 \rx.   \label{psg_sol}
\end{eqnarray}
where $w = u, v$ and $p_1, p_7, p_8, p_9$ are free to take either $0$ or $1$ 
in $\mb{Z}_2$. Details of the derivation can be found in Appendix~\ref{appB}. 
Therefore there are in total 16 states labeled by $p_1, p_7, p_8$ and $p_9$.
Specifically, the state is called $\mathbb{Z}$2A$p_7p_8p_9$ states when $p_1=0$, and 
$\mathbb{Z}$2B$p_7p_8p_9$ states when $p_1 = 1$. This $p_1$ variable is proportional 
to the magnetic flux $p_1 \pi$ through each unit cell felt by the spinon. 
It signifies the fractionalization of translation symmetry 
to be discussed in Sec.~\ref{sec5}.

With the PSG solutions in Tab.~\ref{stab1}, we obtain the mean-field 
Hamiltonians for Schwinger bosons in Appendix~\ref{appC}. The simplified 
results are summarized in Tab.~\ref{tab1}. Due to constraints from the 
PSG, the hermiticity of the Hamiltonian, and time-reversal symmetry, 
some of the coefficients are fixed to be $0$.

\begin{table}
\renewcommand{\arraystretch}{1.2}
\begin{tabular}{c|m{1cm} m{1cm} m{1cm} m{1cm} m{1cm}}
\hline\hline
    $\mathbb{Z}_2$ QSL & $u_s^A$ & $u_a^A$ & $u_s^B$ & $u_a^B$\\\hline
    $\mathbb{Z}$2A000 &  $\neq 0$    & $\neq 0$     & $0$          & $0$\\
    $\mathbb{Z}$2A001 &  $0$         & $\neq 0$     & $\neq 0$     & $0$\\
    $\mathbb{Z}$2A010 &  $\neq 0$    & $\neq 0$     & $\neq 0$     & $0$\\
    $\mathbb{Z}$2A011 &  $0$         & $\neq 0$     & $0$          & $0$\\
    $\mathbb{Z}$2A100 &  $\neq 0$    & $\neq 0$     & $\neq 0$     & $\neq 0$\\
    $\mathbb{Z}$2A101 &  $0$         & $\neq 0$     & $0$          & $\neq 0$\\
    $\mathbb{Z}$2A110 &  $\neq 0$    & $\neq 0$     & $0$          & $\neq 0$\\
    $\mathbb{Z}$2A111 &  $0$         & $\neq 0$     & $\neq 0$     & $\neq 0$\\
    $\mathbb{Z}$2B000 &  $\neq 0$    & $\neq 0$     & $0$          & $0$\\
    $\mathbb{Z}$2B001 &  $0$         & $\neq 0$     & $\neq 0$     & $0$\\
    $\mathbb{Z}$2B010 &  $\neq 0$    & $\neq 0$     & $\neq 0$     & $0$\\
    $\mathbb{Z}$2B011 &  $0$         & $\neq 0$     & $0$          & $0$\\
    $\mathbb{Z}$2B100 &  $\neq 0$    & $\neq 0$     & $\neq 0$     & $\neq 0$\\
    $\mathbb{Z}$2B101 &  $0$         & $\neq 0$     & $0$          & $\neq 0$\\
    $\mathbb{Z}$2B110 &  $\neq 0$    & $\neq 0$     & $0$          & $\neq 0$\\
    $\mathbb{Z}$2B111 &  $0$         & $\neq 0$     & $\neq 0$     & $\neq 0$\\
    \hline\hline
\end{tabular}
\caption{A simplified list of coefficients in the mean-field Hamiltonians 
of each class of $\mb{Z}_2$ QSLs. In the list, $u^A_{s/a}$ stands 
for coefficients for spin-preserving/spin-flipping spinon hopping terms, 
and $u^B_{s/a}$ stands for coefficients for spin-preserving/spin-flipping 
spinon pairing terms. The list emphasizes the vanishing parameters; 
for a complete list, see Tab.~\ref{stab2}.}
\label{tab1}
\end{table}

\begin{figure*}[t]
\centering
{
\includegraphics[width=4.4cm, height=4.4cm]{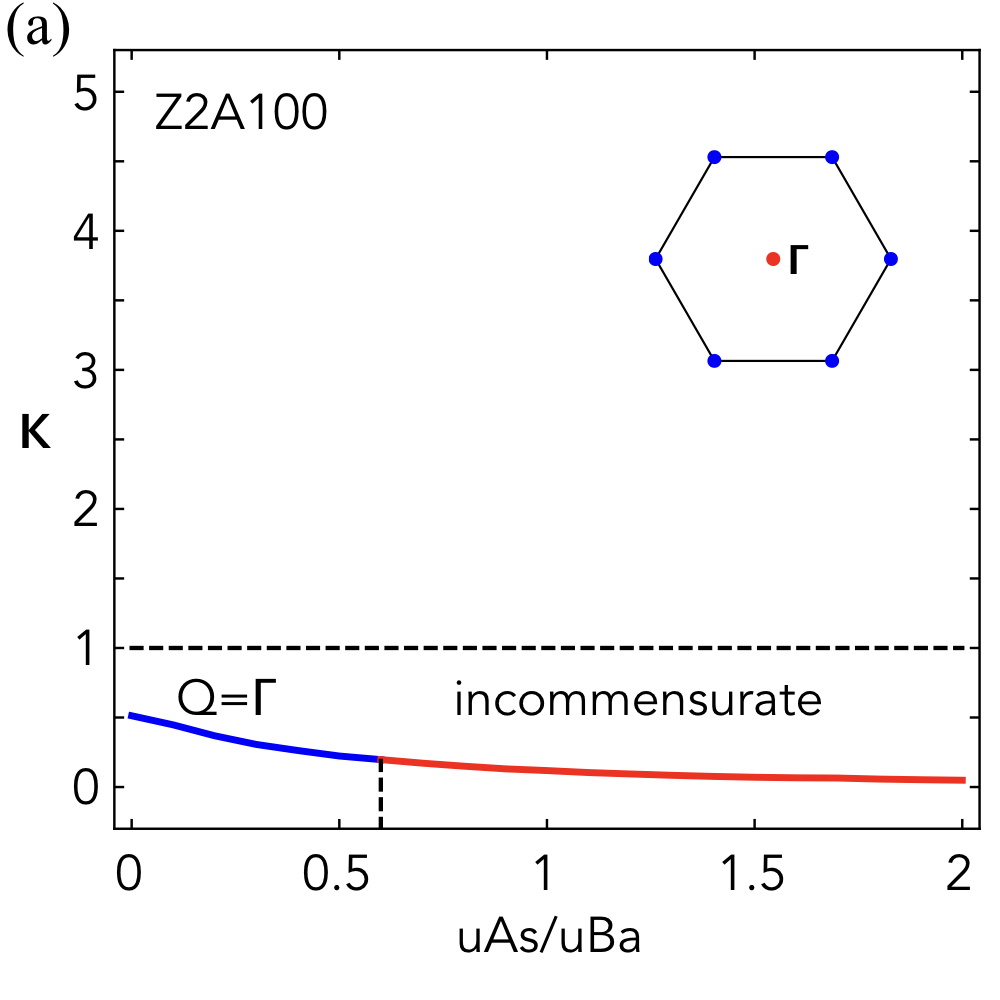}
\includegraphics[width=4.4cm, height=4.4cm]{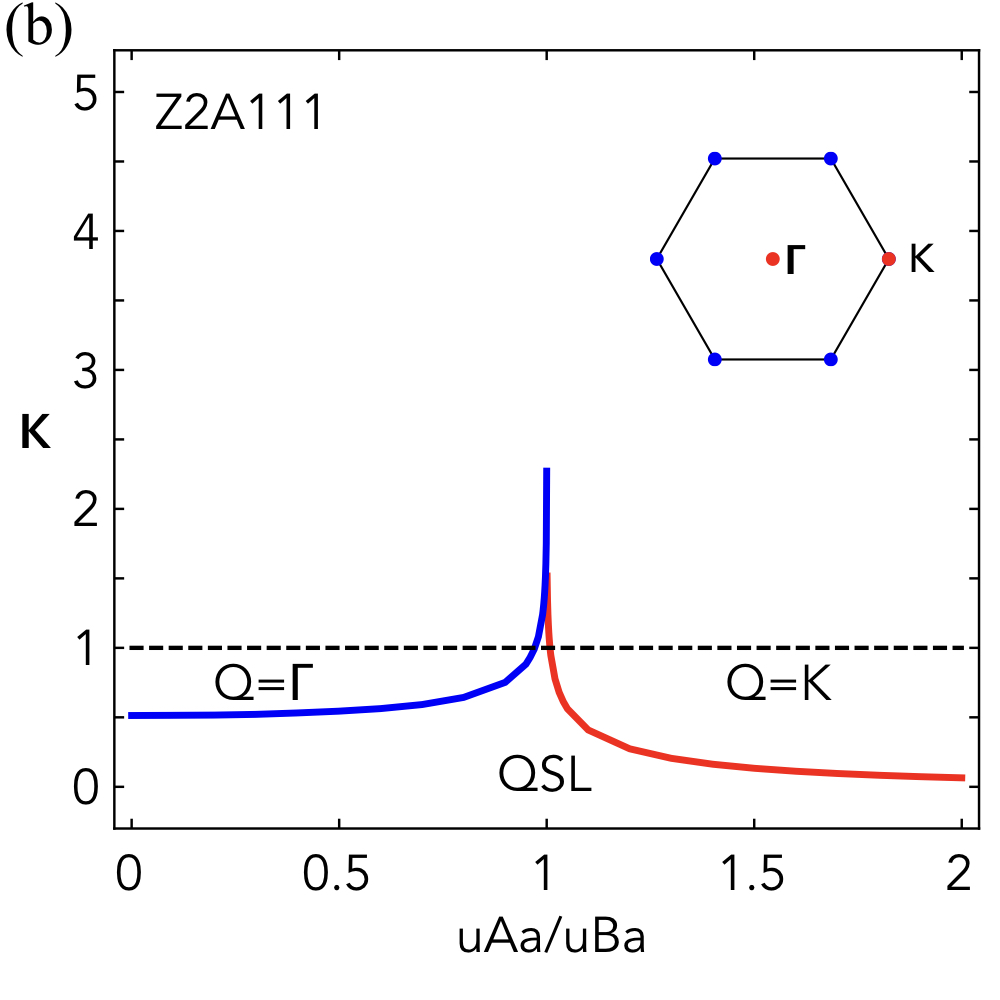}
\includegraphics[width=4.4cm, height=4.4cm]{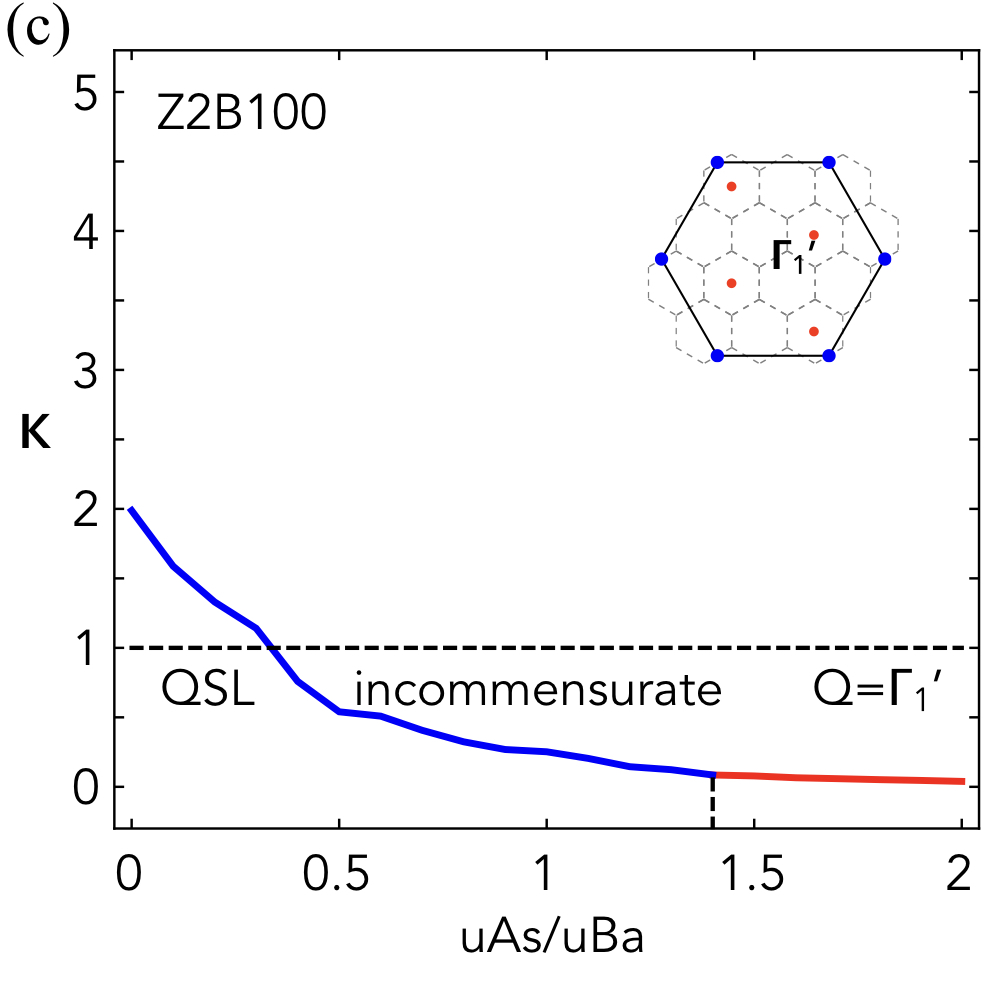}
\includegraphics[width=4.4cm, height=4.4cm]{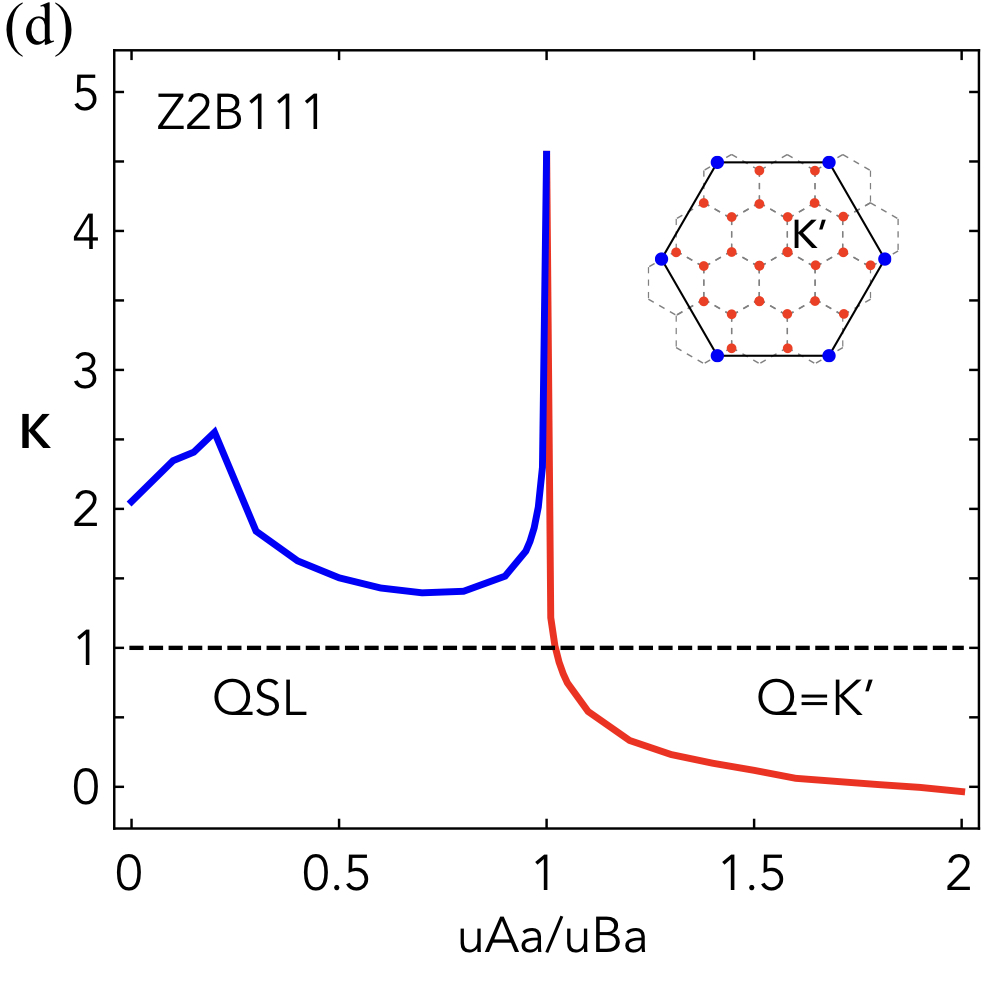}
}
\caption{The phase diagrams for representative mean-field Hamiltonians. 
Here $\kappa$ is the average boson density, defined to be 
$\sum_{i,\alpha} \langle {b^\dg_{i\alpha} b_{i\alpha}^{ } } \rangle / {N_{\rm site}} $, 
and ${\bf Q}$ is the position in Brillouin zone of the spinon band minimum. 
In (a) and (c), we choose ${u^A_a/u^A_s = 0.6}$ and ${u^B_s = 0}$, and in (b) 
and (d) we choose ${u^B_s = 0}$. The solid line marks the phase boundary 
between magnetic ordered state (above solid line) and the $\mathbb{Z}_2$ 
QSL states (below solid line). Here we use different colors for 
solid lines to indicate different ordered states above the solid lines. 
The choice of the momenta can be found in Appendix~\ref{appA}.}
\label{fig2}
\end{figure*}

The classification of $\mb{Z}_2$ QSLs incorporate a wide range of phases 
(at least one for each class) and encode different types of interactions. 
This is particularly relevant to Kitaev materials, where interactions 
beyond the Kitaev model compete with the Kitaev term. These interactions 
can drive the system away from the Kitaev spin liquid state into other 
$\mb{Z}_2$ QSLs, or even destablize the spin liquid and introduce 
a magnetic order. It is therefore desirable to investigate the phase diagram 
for the $\mb{Z}_2$ QSL states in our classification and determine 
the ranges of the parameters that support a QSL phase. We can further 
predict their proximate magnetic orders that can be directly compared with experiments.

The magnetic order out of the $\mb{Z}_2$ QSLs can be understood in the following manner. 
In the $\mb{Z}_2$ QSL phases, the spinons are fully gapped, and the system are absent from 
developing long-range order. However, as we have mentioned in Sec.~\ref{sec2}, the spinon 
density must satisfy the uniform filling condition
\begin{eqnarray}
\kappa = \langle {b^\dg_{i\alpha} b^\pg_{i\alpha}} \rangle = 1. 
\end{eqnarray}
Such a constraint is met by tuning the chemical potential $\mu$ within the mean field theory. 
At a critical value of $\mu$, the spinon gap will close and the spinons condense at the band 
miminum ${\bf Q}$ with $\vev{b_{{\bf Q}\alpha}}\neq 0$. It will correspondingly give rise to 
a magnetic order or spin density wave with ordering wavevector $2{\bf Q}$ (see Sec.~\ref{sec5B}).

Here we choose four representative classes, 
$\mathbb{Z}$2A100, $\mathbb{Z}$2A111, $\mathbb{Z}$2B100, $\mathbb{Z}$2B111, 
and solve for their mean-field phase diagrams (see Fig.~\ref{fig2}). We found 
that the $\mathbb{Z}$2A111, $\mathbb{Z}$2B100, $\mathbb{Z}$2B111 states all support paramagnetic QSL phases in 
the chosen parameter regime, and all of these QSL states can be driven to 
magnetic order when certain parameters are tuned.

\section{Experimental consequences of $\mathbb{Z}_2$ QSLs}
\label{sec5}

In this section we discuss two experimental consequences of the $\mathbb{Z}_2$ QSLs. 
First, we note that translation symmetry fractionalization in $\mathbb{Z}$2B states will result in 
an enhanced periodicity of the lower edge of the dynamic spin structure factor, 
which serves as a direct spectroscopic probe for the QSLs. Second, 
we study magnetic ordered states adjacent to QSLs via the 
condensation of Schwinger bosons. It turns out that the ordering nature of 
the boson-condensed state are determined by the classes of QSLs. 
Therefore the experimentally measured magnetic ordered states will impose 
restrictions on possible adjacent $\mathbb{Z}_2$ QSLs, which helps 
determine the nature of the experimentally realized spin liquids.

\subsection{Spectroscopic signatures of translational symmetry fractionalization}

A unique feature of QSLs is the emergent fractionalized excitations; in our case, 
these are the gapped spinons or visons. The spinons carry quantum numbers that are 
fractions of a physical spin. This fact prevents spinons from being directly probed, 
since any local observable is necessarily with integer quantum number, 
and the observable necessarily adopts a ``convoluted'' form in terms of 
spinon variables. In inelastic neutron scattering experiments, one neutron 
flip event creates a spin-1 excitation, and the energy-transfer of the neutron 
is shared between a pair of spin-1/2 spinons,
\begin{eqnarray}
    \bs{q} &=& \bs{k}_1 + \bs{k}_2, \\
    \Omega(\bs{q}) &=& \omega(\bs{k}_1) + \omega(\bs{k}_2).
\end{eqnarray}
 
In the previous section we classified gapped $\mb{Z}_2$ QSLs on the honeycomb lattice, 
each characterized by a projective representation the emergent spinons live in. It was 
realized that the symmetry class of spinons has dramatic effects on the neutron 
spectrum~\cite{EssinHermeleClassification, EssinHermelePeriodicity}. For 
the lattice translation, the relevant quantum number is $p_1$, and 
we find
\begin{eqnarray}
    \phi_{T_1} (x, y, w) &=& 0, \\
    \phi_{T_2} (x, y, w) &=& p_1 \pi x.
\end{eqnarray}
For the $\mathbb{Z}$2B states, $p_1 = 1$, and the PSG elements corresponding to $T_1$ and $T_2$ 
anticommute,
\begin{eqnarray}
    \hat{T}_1 \hat{T}_2 \hat{T}^{-1}_1 \hat{T}^{-1}_2 = -1.
\end{eqnarray}
where $\hat{T}_1$ and $\hat{T}_2$ act on the spinon degrees of freedom instead on the spins. 
As a consequence, the periodicity of the lower excitation edge of the dynamic spin structure 
factor defined by
\begin{eqnarray}
    {\rm edge}(\bs q) = \min_{\bs{k}} \lz \omega (\bs{k}) + \omega (\bs{q}-\bs{k}) \rz
\end{eqnarray}
is doubled (see Appendix~\ref{appC}),

For the $\mathbb{Z}$2A states, the lower excitation edge should have the 
usual periodicity of $2\pi$ in both directions of Brillouin zone basis.

\begin{figure}[t]
\centering
{
\includegraphics[width=4.2cm]{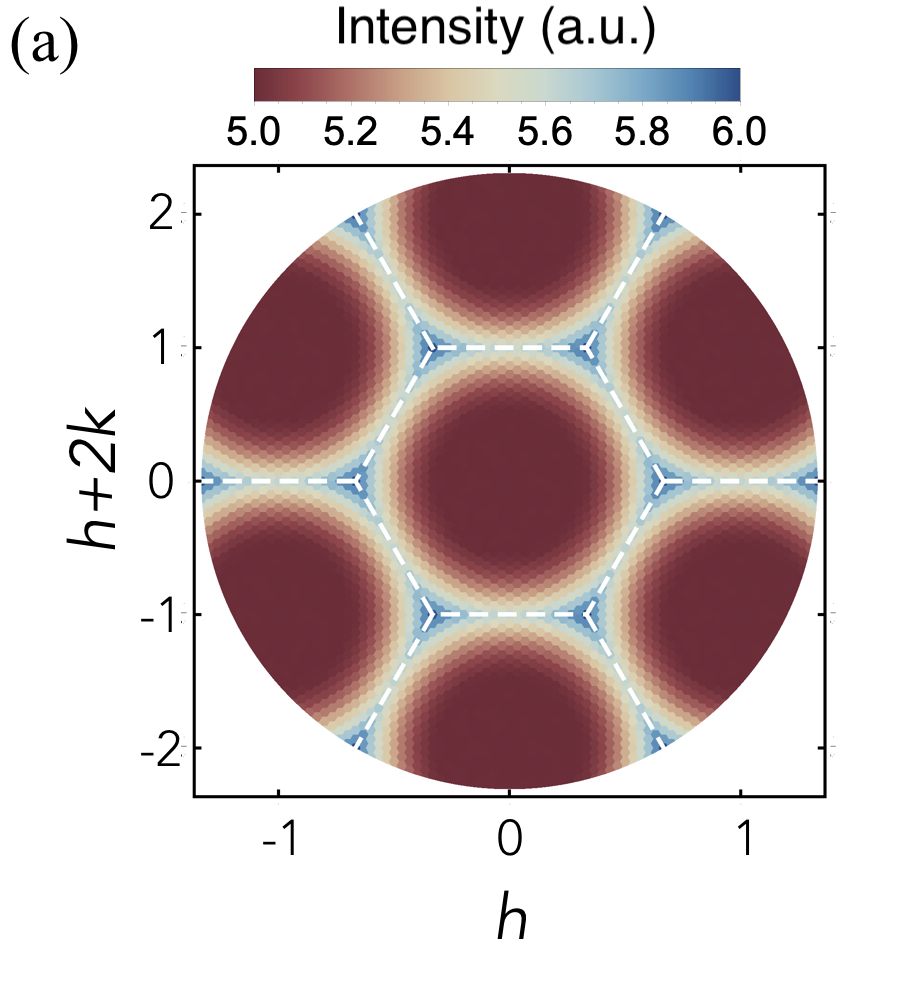}
\includegraphics[width=4.2cm]{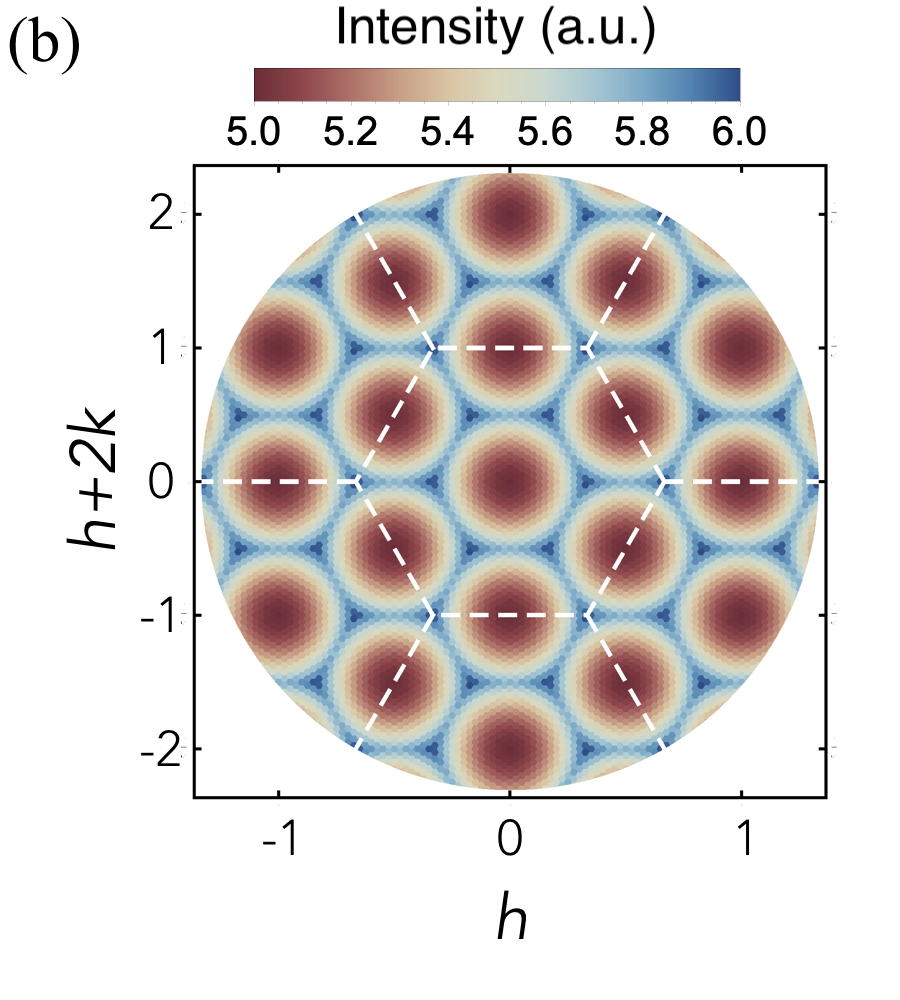}
}
\caption{(Color online.) Intensity plot of lower excitation edges of 
$\mathcal{S}({\bf q},\omega)$ for the (a) $\mathbb{Z}$2A100 and (b) $\mathbb{Z}$2B100 states. 
We have chosen ${u^A_{s} = 2}$, ${u^A_{a} = 1.2}$, ${u^B_{s} = 0}$, ${u^B_{a} = 1}$ 
(see Tab.~\ref{tab1} and Appendix~\ref{appB} for definitions of the parameters). 
The white dashed lines mark the Brillouin zone boundary. 
}
\label{fig3}
\end{figure}

We illustrate the two possible fractionalization patterns in Fig.~\ref{fig3} 
representative $\mathbb{Z}$2A and $\mathbb{Z}$2B states. This pattern is accessible to neutron scattering experiments.

\subsection{Proximate magnetic orders of $\mb{Z}_2$ QSLs}
\label{sec5B}

Besides the symmetry fractionalization in the QSL phases, 
the proximate magnetic orders in the spinon-condensed phases 
provide a complementary description of the system. Instead 
of two-spinon continuum, one expects to see sharp magnon peaks 
in the neutron or the resonant inelastic X-ray scattering data. Therefore, 
the enhanced spectral periodicity in the previous section is no longer 
a relevant description; it is much more feasible to directly 
probe the magnetic order. It would make a strong case for the $\mb{Z}_2$ QSL 
parent state if some of the magnetic orders depicted in Fig.~\ref{fig2} are observed.

In fact, we show here that the proximate magnetic order of the $\mathbb{Z}$2B100 state 
(see Fig.~\ref{fig2}c) has the same ordering wave vector $(\pi, 0)$ as the 
zig-zag order with ordering wave vector observed
in Kitaev materials $\alpha$-RuCl$_3$ and Na$_2$IrO$_3$.

The mean-field Hamiltonian \eqnref{Hmf} of a typical $\mathbb{Z}$2B state in momentum space reads
\begin{eqnarray}
    H = \sum_{\bs{k} \in \frac{1}{2} \text{BZ}} \Psi^\dg_{\bs{k}} (h(\bs{k}) - \mu) \Psi^\pg_{\bs{k}}
\end{eqnarray}
where 
\begin{eqnarray}
\Psi_{\bs{k}} = ( b^\pg_{\bs{k}, w, \alpha, m}, b^\dg_{-\bs{k}, w, \alpha, m} )
\end{eqnarray}
and ${w = u, v}$ labels the $u$ and $v$ sublattices of the honeycomb lattice, 
${\alpha = \ua, \da}$ labels the spin indices, and ${m = 0, 1}$ labels the sites 
in each magnetic unit cell (due to the $\pi$-flux in each of the original unit cell). 
The spectrum has an enhanced periodicity as expected.

\begin{figure*}[tph]
\centering
{
\includegraphics[width=0.235\textwidth]{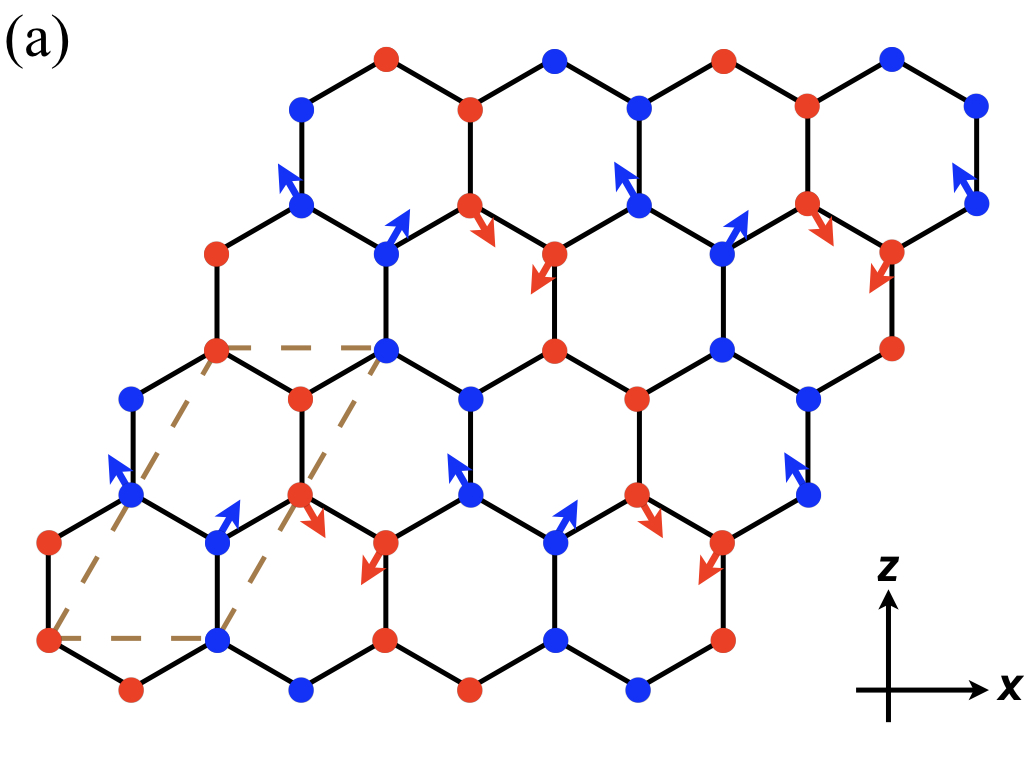}
\includegraphics[width=0.235\textwidth]{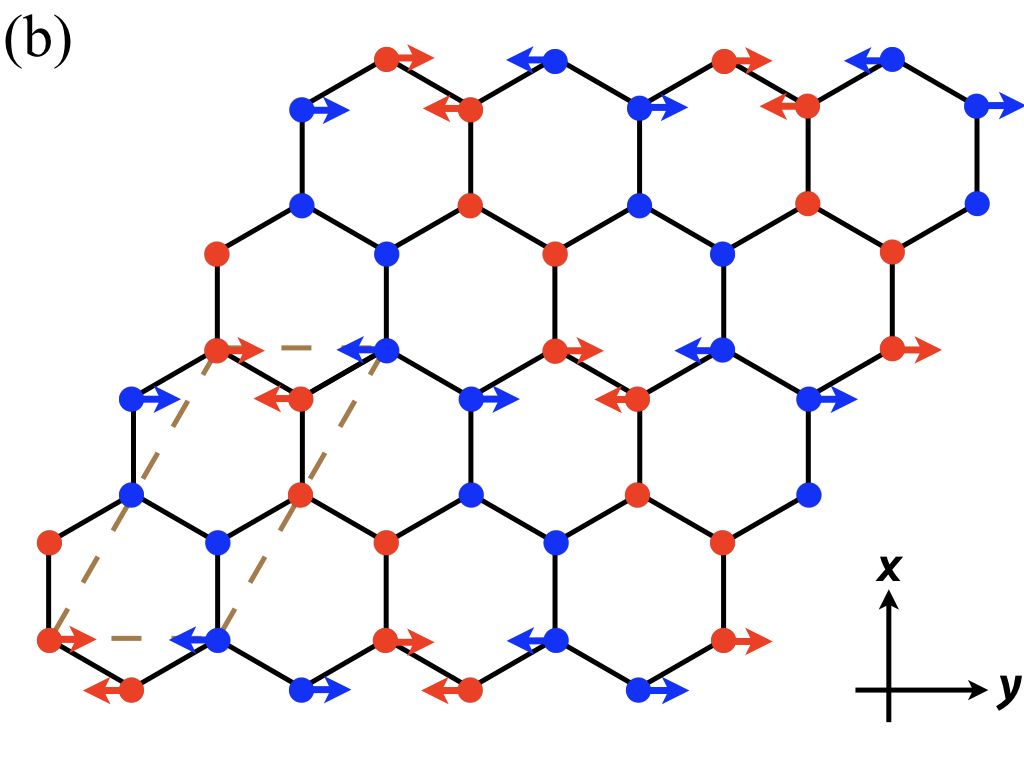}
\includegraphics[width=0.235\textwidth]{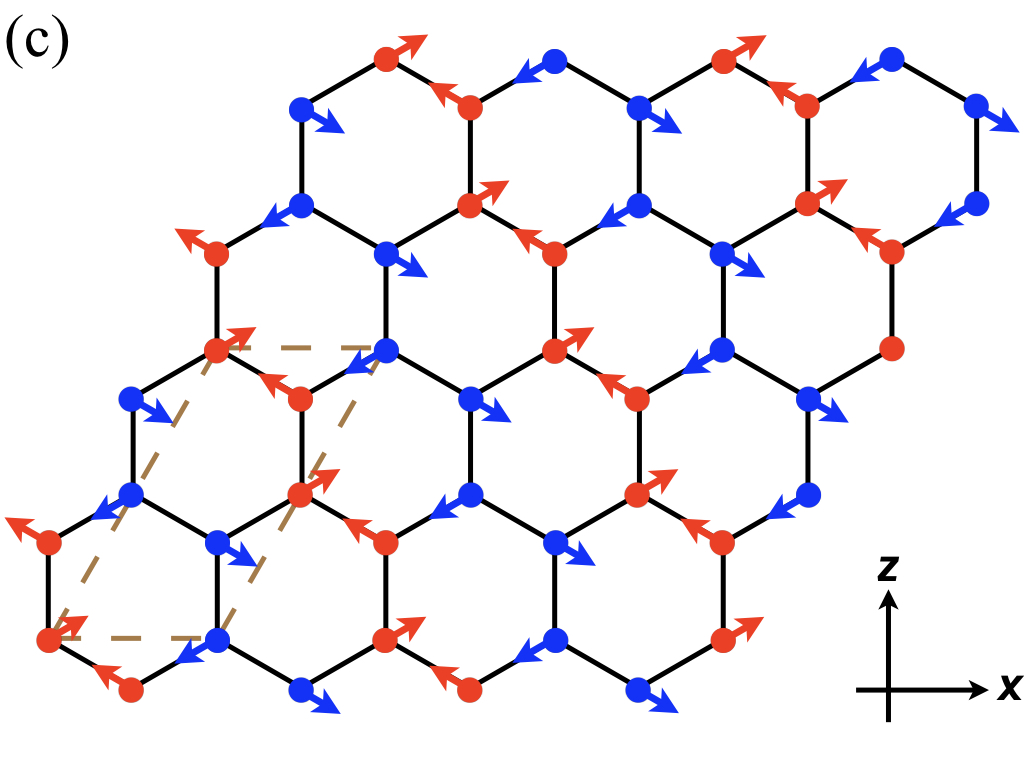}
\includegraphics[width=0.235\textwidth]{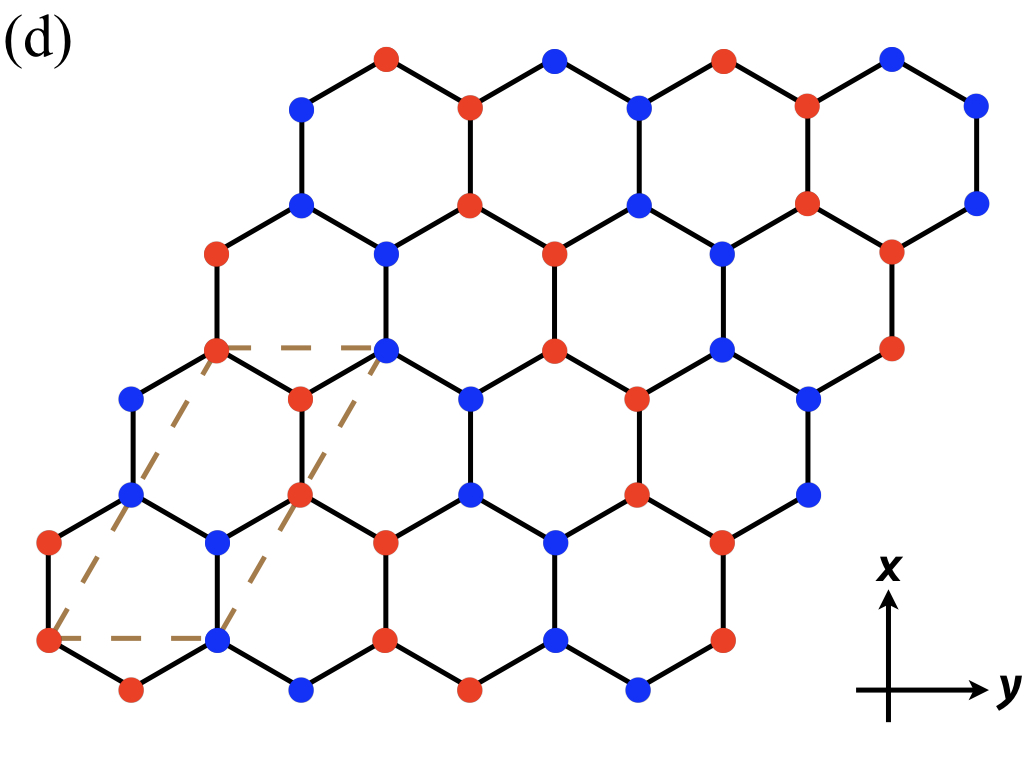}
}
\caption{(Color online.) The magnetic order for a $\mathbb{Z}$2B100 state. We have split the components 
in $x$-$z$ plane and along $y$-direction for clarity. Blue and red sites are 
the antiferromagnetically aligned chains along the direction prependicular to 
$2{\bf Q}$. The gray dashed lines denote the enlarged unit cell. The parameters 
of the Hamiltonian are the same as in Fig.~\ref{fig3}. In (a) and (b), 
we have chosen ${|z_1^{\bf Q}| = |z_2^{\bf Q}|}$, and 
${{\rm arg}\, z_2 - {\rm arg}\, z_1 = \pi/3}$. In (c) and (d), 
we depict the order for ${|z_2^{\bf Q}| = 0}$. Notice that in the 
latter case the magnetic order is completely in the $x$-$z$ plane.
}
\label{fig4}
\end{figure*}

As we see from Fig.~\ref{fig2}c, in a large range of parameters the high symmetry points 
${\pm {\bf Q} = \bs{\Gamma}_1^\p = (\pm \pi/2, 0)}$ are the two independent minimum of the 
spinon band structure in the magnetic Brillouin zone for the $\mathbb{Z}$2B100 state. Moreover, the 
spinons condense at band minima in that regime, and the system is magnetically ordered. 
The corresponding spinon condensate has the following form:
\begin{eqnarray}
   && [
         \langle {b^\pg_{ \bs{r}, u, \ua, 0}} \rangle,
        \langle{b^\dg_{ \bs{r}, u, \ua, 0}}\rangle ,
        \ldots ,
        \langle{b^\pg_{ \bs{r}, v, \da, 1}} \rangle,
        \langle{b^\dg_{ \bs{r}, v, \da, 1}}\rangle
 ]^T \nonumber \\
    &&\quad\quad\quad  = z^{\bf Q}_1 \Psi^{\bf Q}_1 e^{i {\bf Q} \cdot \bs{r}} 
     + z^{\bf -Q}_1 \Psi^{\bf - Q}_1 e^{-i {\bf Q} \cdot \bs{r}} 
\nn && \quad\quad\quad  
    + \,z^{\bf Q}_2 \Psi^{\bf Q}_2 e^{i {\bf Q} \cdot \bs{r}}
    + z^{\bf -Q}_2 \Psi^{\bf -Q}_2 e^{-i {\bf Q} \cdot \bs{r}},
\end{eqnarray}
where $\Psi^{\bf Q}_{1,2}$ and $\Psi^{\bf -Q}_{1,2}$ are eigenvectors of $h(\bs{k})$ 
at $\pm {\bf Q}$ with the lowest energy, respectively.

The choices of the coefficient $z$'s are subject to following constraints:

    1) The condition ${\langle {b^\pg_{\bs{r}, \alpha}} \rangle^\ast 
    = \langle {b^\dg_{\bs{r}, \alpha}} \rangle }$ 
    for all $\bs{r}$ fixes $z^{\bf -Q}_{1,2}$ with respect to $z^{\bf Q}_{1,2}$; 

    2) The boson density 
    ${{ \langle {n_{\bs{r}}} \rangle = \sum_\alpha 
    \langle b^\dg_{ \bs{r}, \alpha}} \rangle	
    \langle b^\pg_{ \bs{r}, \alpha}} \rangle$ 
    should be uniform across the lattice system. 
    This condition will fix ${|z^{\bf Q}_1|^2 + |z^{\bf Q}_2|^2}$.

With the condensate, it is ready to calculate the magnetic order with
\begin{eqnarray}
\langle {\bs{S}_{\bs r}} \rangle = \frac{1}{2} 
\langle b^\dg_{ \bs{r}, \alpha} 
\rangle \bs{\sigma}_{\alpha\beta} 
\langle b^\pg_{\bs{r}, \beta} \rangle.
\end{eqnarray}
We see immediately that the magnetic order has an ordering wave vector of 
${2{\bf Q} = (\pi, 0)}$, consistent with the experimentally observed 
magnetic Bragg peak, and
the magnetic order is controlled by two real parameters while 
the overall phase factor is inessential. 
We have a limited set of free parameters for the magnetic order, 
so the magnetic order would take a rather fixed pattern, 
as illustrated in Fig.~\ref{fig4}. As the zig-zag order, 
the ordering pattern is periodic in the chain direction 
and antiferromagnetic between the neighboring chains.

Although the order differs from the zig-zag or stripe ones, we suspect that 
this is an artifact of the spinon mean-field theory approach. In this framework, 
we are effectively dealing with a theory of free spinons with only nearest neighbor hopping. 
We expect that when further neighbor hoppings and interlayer interactions are taken into account, 
the magnetic order should be closer to reality.
On the other hand, the $\pi$-flux is a robust feature and will survive interactions. 
Consequently, the $2{\bf Q}$ ordering wave vector will exist for a large range of parameters.

To further constrast the $\mathbb{Z}$2B states with the $\mathbb{Z}$2A states, we note that the proximate 
magnetic orders in the phase diagrams of $\mathbb{Z}$2A states in Fig.~\ref{fig2} are either 
incommensurate with the lattice, or have an ordering vector of $2\bs{\Gamma}$ or 
$2{\bf K}$. As a consequence, the resulting magnetic order is either ferromagnetic 
(see Appendix.~\ref{appE}) or an antiferromagnetic order. 
Both are drastically different from the zig-zag order that was observed.

In summary, we have pointed out that the $\mathbb{Z}$2B100 state is likely to be 
the QSL state adjacent to the zig-zag ordered states observed 
in Kitaev materials $\alpha$-RuCl$_3$ and Na$_2$IrO$_3$.

\section{Discussion}
\label{sec6}

The proposed {\sl honeycomb} lattice Kitaev materials are Li$_2$IrO$_3$, Na$_2$IrO$_3$, 
and $\alpha$-RuCl$_3$ with $4d/5d$ magnetic ions. Unfortunately, all three materials develop
long-range magnetic orders, and the relevant magnetic orders were proposed to be the zig-zag 
like with a magnetic unit cell that is twice of the crystal unit cell~\cite{PhysRevB.91.144420,PhysRevB.85.180403,PhysRevLett.110.097204}. For $\alpha$-RuCl$_3$
that is under an active study recently, the magnetic field is found to suppress the magnetism
and possibly generate a QSL state at intermediate magnetic fields. The thermal Hall  
measurement has found a non-vanishing thermal Hall effect that seems to be consistent with 
the prediction from the chiral majorana fermion edge state that is
obtained from the Kitaev spin liquid by the magnetic field~\cite{PhysRevLett.120.217205,PhysRevB.99.085136,Kasahara}. Because of the 
particular experimental setup in the thermal Hall measurements, 
Refs.~\onlinecite{PhysRevLett.121.147201,PhysRevX.8.031032}
carefully considered the effect of the spin-lattice coupling and suggested 
that the quantization of the thermal Hall effect may survive and can actually 
be robust even with the spin-lattice coupling. These results may explain the 
thermal Hall effect in $\alpha$-RuCl$_3$. 
In contrast, our result in this paper is not dealing with the actual spin state in the 
intermediate magnetic fields. Instead, we are interested in the zero-field magnetic state
and try to understand whether the magnetic orders can be thought as the proximate 
magnetic orders of the nearby $\mathbb{Z}_2$ QSLs. Thus, an indirect experimental signature
would be a possible quantum phase transition from the current magnetic orders to the nearly    
$\mathbb{Z}_2$ QSLs. It is not obvious if this transition can be induced by 
the external magnetic field. It is, however, possible that the magnetic field induces 
the magnetic order from the $\mathbb{Z}_2$ QSLs via the spinon condensation where the 
magnetic field suppresses the spinon band gap. 

On the other hand, a recent theoretical development~\cite{PhysRevB.98.184403} 
has extended the Schwinger boson construction to understand the dynamical properties 
of the magnetically ordered state that is obtained by condensing the bosonic spinons. 
Ref.~\onlinecite{PhysRevB.98.184403} applies this theory to study the dynamical properties
of the triangular lattice Heisenberg model, despite this model supports the well-known
120-degree magnetic order. Their results suggested that the Schwinger boson approach 
can be an adequate starting point for describing the excitation spectrum of 
some magnetically ordered compounds that are near the quantum melting point separating 
this ordered phase from the proximate QSL. 
In $\alpha$-RuCl$_3$, the ordered moment is only 
about 1/3 of the full magnetic moment in the paramagnetic phase~\cite{Banerjee1055}.
Thus, it is natural and interesting to see whether Ref.~\onlinecite{PhysRevB.98.184403}'s
approach can be adapted to provide a new understanding of the spin dynamics inside 
the magnetic ordered state of $\alpha$-RuCl$_3$ rather than making connection to 
the Kitaev spin liquid.

Quite recently, the pressurized $\alpha$-RuCl$_3$ has been studied experimentally~\cite{PhysRevB.97.245149}, 
as well as other strain effect experiments have been performed. We focus our discussion
on the pressurized experiments~\cite{PhysRevB.97.245149}. It is found that, 
above a critical pressure, the antiferromagnetic order in $\alpha$-RuCl$_3$ 
disappears and a possible QSL state appears. At even higher pressures, the system 
experiences a resistance drop by several orders in magnitude. 
This was interpreted as the softening or the closing of the charge gap. 
At the mean time, the magnetoresistivity in this range of pressure 
remain insensitive to the magnetic field up to 7T.  There are 
several puzzles associated with this pressurized experiment. 
What is the nature of the disordered state when the magnetic order diappears? 
What is the nature of the disordered state with a significantly reduced resistance
in the high pressure regime? 
What do the spin degrees of freedom do in this high pressure regime? 
The experimental information is quite limited to address these questions. 
However, here we would like make a bold suggestion. First, we discuss 
the possibility that 
the disordered state can be a QSL state. The absence of the phase transition
in the heat capacity measurement down to 4K suggests that the candidate QSL 
cannot be a symmetry broken state such as the time reversal symmetry broken
chiral spin liquid. From the robustness of a phase in a large range of pressures,
the candidate state may be a $\mathbb{Z}_2$ QSL, and this $\mathbb{Z}_2$
topological order would survive even to the pressure when the 
charge gap is suppressed. In fact, Ref.~\onlinecite{PhysRevB.97.245149} 
has attributed the insensitivity of the 
magnetoresistance to the magnetic field to the dominance of the spin 
energy scale. In the future experiments, it will be interesting to 
perform an inelastic neutron scattering measurement to check if the spinon 
continuum shows a spectral periodicity enhancement. In addition, doping the 
pressurized materials and examining the possibility of superconductivity 
or non-Fermi liquid behaviors can be quite interesting too. 
It is interesting to notice that doping the spin-orbit-coupled Mott insulators
such as RuCl$_3$, iridates, or any others with spin-orbit-entangled 
local moments beyond the $4d^5/5d^5$ ${j=1/2}$ moments would 
necessarily experience an electron-hole doping asymmetry. This doping 
asymmetry arises from the distinct spin-orbital reconstruction/entanglement of 
the different electron occupation configurations from electron and hole doping. 
We will elaborate this general point in a later paper. 
Furthermore, if the pressurized sample develops a dimerized state, a natural question 
would be the nature of the phase transtion between the zig-zag magnetic order
and the dimerization. Could this transition be a deconfined quantum criticality 
that is very much like the N\'{e}el-VBS transition proposed for the square lattice 
antiferromagnets~\cite{dqcp}? As the pressure can be tuned continuously, this 
question could be addressed experimentally by tuning the pressure to the transition 
point in the future. The other question would be whether the dimerized state can 
be obtained by condensing visons from the same $\mathbb{Z}_2$ QSL that gives the 
zig-zag magnetic order. These two questions can be pushed forward when 
more experimental results are available.

To summarize, in this paper we have carefully classified the possible $\mathbb{Z}_2$ QSLs 
and studied the experimental signatures such as the proximate magnetic orders, symmetry 
fractionalization of the spinons, and the structure of the spinon continuum. Our results 
provide a rather different perspective from the existing thoughts on these Kitaev materials.

\section{Acknowledgments}

We thank Dr. Jiawei Mei and Dr. Khaliullin for telling us the possibility 
of spin dimerization in the pressurized sample. 
This work is supported by the ministry of science and technology of China 
with Grant No.2016YFA0301001, 2016YFA0300500, 2018YFGH000095.

\appendix
\section{The coordinate system and space group}
\label{appA}

The honeycomb lattice is illustrated in Fig.~\ref{fig1} of the main text. 
We choose the basis vectors to be
\begin{eqnarray}
    \bs{a}_1 = (1,0), \quad
    \bs{a}_2 &=& ( \frac{1}{2}, \frac{\sqrt{3}}{2} ).
\end{eqnarray}
The lattice sites are labeled by $(x, y, w)$, where $w = u, v$ is the 
sublattice index. The position of the site $(x, y, w)$ is
\begin{eqnarray}
    \bs{r} (x, y, w) = \begin{cases}
                x \bs{a}_1 + y \bs{a}_2, \quad\quad\quad\quad\quad\quad \text{if\ } w = u,\\
                x \bs{a}_1 + y \bs{a}_2 + \lx 0, \frac{1}{\sqrt{3}} \rx, \quad \, \text{if\ } w = v.\\
            \end{cases}
\end{eqnarray}

 All momenta vectors are represented in the $\{\bs{v}_1, \bs{v}_2\}$ basis, where
\begin{eqnarray}
    \bs{v}_1 = ( 1, -\frac{1}{\sqrt{3}} ), \quad \bs{v}_2 = ( 0, \frac{2}{\sqrt{3}} ),
\end{eqnarray}
so that $\bs{a}_i \cdot \bs{v}_j = \delta_{ij}$.
Therefore, the basis vectors of the Brillouin zone has the following forms,
\begin{eqnarray}
    \bs{b}_1 = (2\pi, 0),\quad \bs{b}_2 = (0, 2\pi),
\end{eqnarray}
and 
\begin{eqnarray}
    \bs{\Gamma} &=& (0, 0),\\
    {\bf M}  &=& (\pi, 0) \text{ or } (0, \pi) \text{ or } (\pi, \pi) \\
    {\bf K} &=&  ( \frac{4\pi}{3}, \frac{2\pi}{3} ) \text{ or } ( \frac{2\pi}{3}, \frac{4\pi}{3} ).
\end{eqnarray}
We define additional high-symmetry points in the Brillouin zone,
\begin{eqnarray}
    \bs{\Gamma}^\p &=& ( \frac{\pi}{2}, \frac{\pi}{2} ) 
    + a ( \frac{2\pi}{3}, \frac{\pi}{3} ) 
    +  b ( \frac{\pi}{3}, \frac{2\pi}{3} ),
\end{eqnarray}
where $a, b \in \mb{Z}$. $\bs{\Gamma}_1^\p$ corresponds to those 
with $a-b = 0 (\text{mod }3)$, and $\bs{\Gamma}_2^\p$ those with 
$a-b \neq 0 (\text{mod } 3)$.
Finally,
\begin{eqnarray}
    {\bf K}^\p = \bs{\Gamma}^\p + ( 0, \frac{\pi}{3} ) 
    \text{ or } 
    {\bf K}^\p = \bs{\Gamma}^\p + ( \frac{\pi}{3}, \frac{\pi}{3} ).
\end{eqnarray}

The symmetry group of the honeycomb lattice consists of translations $T_1, T_2$, 
a six-fold rotation $C_6$ and a reflection $\sigma$. Explicitly in terms of the 
lattice indices, their actions read
\begin{eqnarray}
   T_1&:& (x,y,w) \to (x+1,y,w),\quad w = u,v \\
   T_2&:& (x,y,w) \to (x,y+1,w),\quad w = u,v \\
   C_6&:&  \begin{cases}
               (x,y,u) \to (-y+1, x+y-1, v),\\
               (x,y,v) \to (-y, x+y, u),
           \end{cases} \\
   \sigma&:& \begin{cases}
               (x,y,u) \to (x+y, -y, v), \\
               (x,y,v) \to (x+y, -y, u).
             \end{cases}
\end{eqnarray}

\section{Algebraic solution of the $\mb{Z}_2$ PSG on honeycomb lattice}
\label{appB}

In this appendix we show classification of algebraic $\mb{Z}_2$ QSLs 
by solving the PSG defined in Sec.~\ref{sec3}.

The space group of the honeycomb lattice and its elements are defined in Sec.~\ref{sec3}. 
Presentations of the space group are
\begin{eqnarray}
   && T_1^{-1} T_2^\pg T_1^\pg T_2^{-1}
    = T_1^{-1} C_6^\pg T_1^\pg T_2^{-1} C_6^{-1}
  = T_2^{-1} C_6^\pg T_1^\pg C_6^{-1}
   \nn
   && \quad = C_6^6=T_1^{-1} \sigma T_1^\pg \sigma^{-1}
    = T_2^{-1} \sigma T_1^\pg T_2^{-1} \sigma^{-1} 
    \nn
    & &\quad = \sigma^2
    = \sigma C_6 \sigma C_6
    = {\bf 1}.
\end{eqnarray}

We will assume the IGG is $\mb{Z}_2$ (see Sec.~\ref{sec3}), 
and assume the generator of the IGG is
\begin{eqnarray}
    b_{j\alpha} \to -b_{j\alpha}, \quad \alpha = \ua, \da,\quad \forall j.
\end{eqnarray}
Elements of the IGG obviously preserves all mean field ansatz; 
therefore, the classification of algebraic spin liquid states 
are determined up to an IGG element.

For each space group element $\mc{O}$, we associate a U(1) phase 
${\mc{G}^\mc{O}_{\mc{O}(i)} = e^{i \phi_{\mc{O}}[\mc{O}(i)]}}$ 
such that the mean field Hamiltonian is invariant under the combined PSG operation,
\begin{eqnarray}
    b^\pg_{i} \to \mc{G}^\mc{O}_{\mc{O}(i)} \, \mc{U}^\pg_\mc{O} \, b^\pg_{\mc{O}(i)}.
\end{eqnarray}
The $\mc{U}$ matrices accounts for the effects of SOC (see their definitions in Sec.~\ref{sec3}).

Before solving for the PSG, we consider the effect of a pure gauge transformation 
${\mc{G}: b_{i\alpha} \to e^{i\phi_\mc{G}(i)} b_{i\alpha}}$ on the U(1) phases 
$\mc{G}^\mc{O}$ associated to each group element. The symmetry operation on 
the gauge transformed boson reads 
${\mc{G} \mc{G}^\mc{O} \mc{U}_\mc{O} \mc{O} \mc{G}^{-1} 
= \mc{G} \mc{G}^\mc{O} \mc{U}_\mc{O} \mc{O} \mc{G}^{-1} \mc{O}^{-1} \mc{O}}$.
Since $\mc{U}_\mc{O}$ commutes with $\mc{G}$, $\mc{G}^\mc{O}$, 
and $\mc{O}$, $\mc{U}_\mc{O}$ cancel on both sides. 
Therefore $\mc{G}^\mc{O}$ shoud be replaced by 
$\mc{G} \mc{G}^\mc{O} \mc{O} \mc{G}^{-1} \mc{O}^{-1}$, 
or~\cite{WangAshvin}
\begin{eqnarray}
    \phi_\mc{O}(i) \to \phi_\mc{G}(i) + \phi_\mc{O}(i) - \phi_\mc{G}(\mc{O}^{-1}(i)).
\end{eqnarray}
Using the gauge freedom one can always assume (open boundary condition)
\begin{eqnarray}
    \phi_{T_1}(x,y,w)=0, \quad \phi_{T_2}(x=0,y,w)=0. 
    \label{eqn17}
\end{eqnarray}
For the honeycomb lattice, this can be achieved by solving equations
\begin{eqnarray}
    \phi_\mc{G}(x,y,w)-\phi_\mc{G}(x-1,y,w)+\phi_{T_1}(x,y,w)=0, \label{eqn18} \\
    \phi_\mc{G}(0,y,w)-\phi_\mc{G}(0,y-1,w)+\phi_{T_2}(0,y,w)=0. \label{eqn19}
\end{eqnarray}

For simplicity of notations we define $\Delta_1 f(x,y)=f(x+1,y)-f(x,y)$ 
and $\Delta_2 f(x,y)=f(x,y+1)-f(x,y)$.

The identity ${T_1^{-1} T_2^\pg T_1^\pg T_2^{-1} = {\bf 1}}$ 
translates into the following equation of PSG elements,
\begin{eqnarray}
    (\mc{G}^{T_1} \mc{U}_{T_1} T_1)^{-1} (\mc{G}^{T_2} \mc{U}_{T_2} T_2) (\mc{G}^{T_1} \mc{U}_{T_1} T_1) (\mc{G}^{T_2} \mc{U}_{T_2} T_2)^{-1} = \pm {\bf 1}, \nn
\end{eqnarray}
where the RHS is an element of the IGG. In terms of phases,
\begin{eqnarray}
    &&-\phi_{T_1} [T_1({\bf r})] + \phi_{T_2}[T_1({\bf r})] + \phi_{T_1}[T_1 T_2^{-1} ({\bf r})] - \phi_{T_2} [{\bf r}] \nn
    &&\quad \quad\quad\quad \quad\quad\quad = \Delta_1 \phi_{T_2} (x,y,w) \nn
    &&\quad \quad\quad\quad \quad\quad\quad = p_1 \pi,
\end{eqnarray}
where ${{\bf r} = (x,y,w), p_1 \in \mb{Z}_2}$, and we have adopted 
assumption in \eqnref{eqn17}. Again from \eqnref{eqn17}, we see that
\begin{eqnarray}
    \phi_{T_2}(x,y,w) = p_1 \pi x.
\end{eqnarray}
In other words, the flux in one elementary hexagon is $p_1 \pi$.

Similarly, from ${T_1^{-1} C_6^\pg T_1^\pg T_2^{-1} C_6^{-1} = T_2^{-1} C_6^\pg T_1^\pg C_6^{-1} = {\bf 1}}$ 
we have
\begin{eqnarray}
  &&  (\mc{G}^{T_1} \mc{U}_{T_1} T_1)^{-1} (\mc{G}^{C_6} \mc{U}_{C_6} C_6) (\mc{G}^{T_1} \mc{U}_{T_1} T_1) \nn
  && \quad\quad \quad\quad (\mc{G}^{T_2} \mc{U}_{T_2} T_2)^{-1}  (\mc{G}^{C_6} \mc{U}_{C_6} C_6)^{-1} =\pm {\bf 1}, \\
  &&  (\mc{G}^{T_2} \mc{U}_{T_2} T_2)^{-1} (\mc{G}^{C_6} \mc{U}_{C_6} C_6) \nn
  && \quad\quad\quad\quad 
    (\mc{G}^{T_1} \mc{U}_{T_1} T_1)  (\mc{G}^{C_6} \mc{U}_{C_6} C_6)^{-1} =\pm {\bf 1}.
\end{eqnarray}
The $\mc{U}_\mc{O}$'s cancel, and the solution for the U(1) phases is
\begin{eqnarray}
    \Delta_1 \phi_{C_6}(x,y,w) &=& p_1 \pi(x+y) + p_2 \pi, \\
    \Delta_2 \phi_{C_6}(x,y,w) &=& p_1 \pi x + p_3 \pi.
\end{eqnarray}

Performing pure gauge transformations, we may further assume~\cite{FaWang}
\begin{eqnarray}
    p_2 = p_3 = 0, \quad \phi_{C_6} (0,0,u) = \phi_{C_6} (0,0,v). \label{eqn46}
\end{eqnarray}
The solution for $\phi_{C_6}$ reads
\begin{eqnarray}
    \phi_{C_6}(x,y,w) &=& \phi_{C_6}(0,0,w) \nn
    &&  + p_1 \pi \frac{x(x+2y-1)}{2}.
\end{eqnarray}

In the PSG formulation of the group relation 
$ T_1^{-1} \sigma T_1^\pg \sigma^{-1} =  T_2^{-1} \sigma T_1^\pg T_2^{-1} \sigma^{-1} = {\bf 1}$, the $\mc{U}_{\sigma}$'s again cancel. Thus we have
\begin{eqnarray}
    \Delta_1 \phi_\sigma (x,y,w) &=& p_4 \pi, \\
    \Delta_2 \phi_\sigma (x,y,w) &=& p_1 \pi y + p_5 \pi.
\end{eqnarray}
The solution for $\phi_{\sigma}$ is
\begin{eqnarray}
    \phi_\sigma(x,y,w) &=& \phi_\sigma(0,0,w) 
                        +  \frac{1}{2} p_1 \pi y(y-1) \nn
    && 
    \quad\quad + p_4 \pi x + p_5 \pi y.
\end{eqnarray}

From $C_6^6 = {\bf 1}$ we have
\begin{eqnarray}
      (\mc{G}^{C_6} \mc{U}_{C_6} C_6)^6
    = (\mc{U}_{C_6})^6 (\mc{G}^{C_6} C_6)^6
    = \pm {\bf 1},
\end{eqnarray}
since $\mc{U}_{C_6}$ acts only on the spin indices and commutes with 
$\mc{G}^{C_6}$ and $C_6$. Since $(\mc{U}_{C_6})^6 = -{\bf 1}$, the 
above equation simplifies to $(\mc{G}^{C_6} C_6)^6 = \pm {\bf 1}$, giving
\begin{eqnarray}
    && 3[\phi_{C_6}(0,0,u)+\phi_{C_6}(0,0,v)] 
\nn    
    && \quad\quad\quad\quad = (p_1+p_2)\pi + p_6 \pi. 
    \label{eqn33} 
\end{eqnarray}

For ${\sigma^2 = {\bf 1}}$ we have
\begin{eqnarray}
     (\mc{G}^{\sigma} \mc{U}_{\sigma} \sigma)^2
    &=& (\mc{U}_{\sigma})^2 (\mc{G}^{\sigma} \sigma)^2 \nn
    &=& - (\mc{G}^{\sigma} \sigma)^2
    = \pm {\bf 1},
\end{eqnarray}
where $\mc{U}_{\sigma}$ commutes with the rest, and ${(\mc{U}_{\sigma})^2 = -{\bf 1}}$. 
This results in the constraint
\begin{eqnarray}
   && \phi_\sigma(0,0,u) + \phi_\sigma(0,0,v) \nn
   &&\quad\quad\quad\quad\quad = (p_1 y^2 + p_4 y + p_7) \pi  .
\end{eqnarray}
We see immediately ${p_1 = p_4}$ by comparing ${y=0}$ and ${y=1}$ in this equation.

From $\sigma C_6 \sigma C_6$ we have
\begin{eqnarray}
    && (\mc{G}^{\sigma} \mc{U}_{\sigma} \sigma) (\mc{G}^{C_6} \mc{U}_{C_6} C_6) (\mc{G}^{\sigma} \mc{U}_{\sigma} \sigma) (\mc{G}^{C_6} \mc{U}_{C_6} C_6) \nn
    &=& (\mc{U}_{\sigma} \mc{U}_{C_6} \mc{U}_{\sigma} \mc{U}_{C_6}) (\mc{G}^{\sigma} \sigma) (\mc{G}^{C_6} C_6) (\mc{G}^{\sigma} \sigma) (\mc{G}^{C_6} C_6) \nn
    &=& - (\mc{G}^{\sigma} \sigma) (\mc{G}^{C_6} C_6) (\mc{G}^{\sigma} \sigma) (\mc{G}^{C_6} C_6) \nn
    &=& \pm {\bf 1}.
\end{eqnarray}
Therefore we have
\begin{eqnarray}
    2\phi_\sigma(0,0,v)+2\phi_{C_6}(0,0,u) &=& 2\phi_\sigma(0,0,u)+2\phi_{C_6}(0,0,v)\nn &=& p_8 \pi.
\end{eqnarray}
and $p_1 = p_5$. Due to \eqnref{eqn46}, we see that $2\phi_\sigma(0,0,v) = 2\phi_\sigma(0,0,u)$, giving
\begin{eqnarray}
    \phi_\sigma(0,0,u)-\phi_\sigma(0,0,v) = p_9 \pi.
\end{eqnarray}

$\phi_{C_6} (0,0,w)$ and $\phi_\sigma (0,0,w)$ can be solved,
\begin{eqnarray}
    \phi_\sigma(0,0,u) = (p_7 + p_9)\pi/2\ {\rm mod}\ 2\pi, \\
    \phi_\sigma(0,0,v) = (p_7 - p_9)\pi/2\ {\rm mod}\ 2\pi, \\
    \phi_{C_6}(0,0,w) = (p_7 + p_8 + p_9)\pi/2\ {\rm mod}\ 2\pi,
\end{eqnarray}
and $p_6 = p_1 + p_7 + p_8 + p_9$ from \eqnref{eqn33}.

Summarizing, the solutions of the PSG are
\begin{eqnarray}
  && \phi_{T_1}(x,y,w) = 0,   \\
  && \phi_{T_2}(x,y,w) = p_1 \pi x,   \\
  && \phi_{C_6}(x,y,w) = \frac{\pi}{2} \big[ p_1 x (x+2y-1) \nonumber \\
  &&  \quad\quad\quad\quad\quad  \quad\quad\quad\quad\quad+ p_7 + p_8 + p_9\big],   \\
  && \phi_{\sigma}(x,y,u) = \frac{\pi}{2} \big[ 2p_1x+p_1y(y+1)+p_7+p_9 \big],   \\
  && \phi_{\sigma}(x,y,v) = \frac{\pi}{2} \big[ 2p_1x+p_1y(y+1)+p_7-p_9 \big].  \label{psg_sol2}
\end{eqnarray}
where $w = u, v$ and $p_1, p_7, p_8, p_9$ are free to take either $0$ or $1$ in $\mb{Z}_2$. 
There are in total 16 possible classes of QSLs; the mean field ansatz would further 
constrain the number of free parameters.
The respective gauge transformations for the 16 $\mb{Z}_2$ QSLs are summarized in Tab.~\ref{stab1}.

\section{Nearest neighbor mean field ansatz of the $\mb{Z}_2$ PSG}
\label{appC}

\begin{center}
\begin{table*}[htp]
\renewcommand{\arraystretch}{1.5}
\begin{tabular}{c|c c c  }
\hline\hline
    Bond & $(x,y,u)$-$(x,y,v)$ & $(x,y,u)$-$(x+1,y-1,v)$ & $(x,y,u)$-$(x,y-1,v)$ \\\hline
    $u^A_{\ua\ua}$ &  $u^A_s$         & $(-)^{p_1(y+1)}u^A_s$          & $u^A_s$         \\
    $u^A_{\da\da}$ &  $(-)^{p_9} u^A_s$         & $(-)^{p_1(y+1)+p_9}u^A_s$  & $(-)^{p_9}u^A_s$       \\
    $u^A_{\ua\da}$ &  $e^{i(p_9/2)\pi}u^A_a$         &  $e^{i(p_9/2+2/3+p_1(y+1))\pi}u^A_a$          & $e^{i(p_9/2-2/3)\pi}u^A_a$         \\
    $u^A_{\da\ua}$ &  $-e^{-i(p_9/2)\pi}u^A_a$         & $-e^{-i(p_9/2+2/3+p_1 (y+1))\pi}u^A_a$         & $-e^{-i(p_9/2-2/3)\pi}u^A_a$      
    \\
    $u^B_{\ua\ua}$ &  $u^B_s$         & $e^{i(-2/3+p_1(y+1))\pi}u^B_s$          & $e^{i(+2/3)\pi}u^B_s$         \\
    $u^B_{\da\da}$ &  $(-1)^{1+p_7}u^B_s$         & $(-1)^{1+p_7} e^{i(+2/3+p_1(y+1))\pi}u^B_s$          & $(-1)^{1+p_7} e^{i(-2/3)\pi}u^B_s$         \\
    $u^B_{\ua\da}$ &  $u^B_a$         & $(-)^{p_1(y+1)}u^B_a$          & $u^B_a$       \\
    $u^B_{\da\ua}$ &  $(-)^{p_7+p_8+p_9}u^B_a$      & $(-)^{p_7+p_8+p_9+p_1 (y+1)}u^B_a$    & $(-)^{p_7+p_8+p_9}u^B_a$      \\
    \hline\hline
        Bond &   $(x,y,v)$-$(x,y,u)$ & $(x,y,v)$-$(x,y+1,u)$ & $(x,y,v)$-$(x-1,y+1,u)$ \\\hline
    $u^A_{\ua\ua}$ &           $u^A_s$       & $u^A_s$   &  $(-)^{p_1 y}u^A_s$ \\
    $u^A_{\da\da}$ &         $(-)^{p_9}u^A_s$       & $(-)^{p_9}u^A_s$   &  $(-)^{p_1 y+p_9}u^A_s$ \\
    $u^A_{\ua\da}$ &       $-e^{i(p_9/2)\pi}u^A_a$        & $-e^{i(p_9/2-2/3)\pi}u^A_a$ & $-e^{i(p_9/2+2/3+p_1 y)\pi}u^A_a$\\
    $u^A_{\da\ua}$ &  $e^{-i(p_9/2)\pi}u^A_a$       & $e^{-i(p_9/2-2/3)\pi}u^A_a$ & $e^{-i(p_9/2+2/3+p_1 y)\pi}u^A_a$\\
    $u^B_{\ua\ua}$ & $u^B_s$       & $e^{i(+2/3)\pi}u^B_s$ & $e^{i(-2/3+p_1 y)\pi}u^B_s$\\
    $u^B_{\da\da}$  & $(-1)^{1+p_7} u^B_s$       & $(-1)^{1+p_7} e^{i(-2/3)\pi}u^B_s$ & $(-1)^{1+p_7} e^{i(+2/3+p_1 y)\pi}u^B_s$\\
    $u^B_{\ua\da}$    & $(-)^{p_7+p_8+p_9}u^B_a$       & $(-)^{p_7+p_8+p_9}u^B_a$ & $(-)^{p_7+p_8+p_9+p_1 y}u^B_a$\\
    $u^B_{\da\ua}$   & $u^B_a$      & $u^B_a$ & $(-)^{p_1 y}u^B_a$\\
    \hline\hline
\end{tabular}
\caption{Spatial patterns of the nearest-neighbor mean-field ansatz. $u^A_s, u^A_a, u^B_a$ and $u^B_s$ are real numbers. The coefficients are subject to constraints from hermiticity and time reversal symmetry.}
\label{stab2}
\end{table*}

\end{center}

In this appendix we present symmetry allowed mean field ansatz up to nearest neighbors.

The algebraic solution of PSG is very general and usually contains many free parameters.
 Certain mean field ansatz will put further constraints on the PSG. In particular, if a non-identity space group element $\mc{O}$ transforms a bond to itself or its inverse, the form of exchange terms on this bond will be constrained.

We first consider the spin-flipping pairing terms ($u^B_a$ terms).
Under the action of $\sigma$,
\begin{eqnarray}
&&    u^B_a b_{(0,0,u) \ua} b_{(0,0,v) \da} \nn
&&\quad\quad\quad    \to e^{i (p_7 + 1) \pi} u^B_a b_{(0,0,u) \ua} b_{(0,0,v) \da}.
\end{eqnarray}
Therefore nonzero $u^B_a$ requires ${p_7 = 1}$. Under $T_1^{-1}C_6^3$,
\begin{eqnarray}
 &&    u^B_a b_{(0,0,u) \ua} b_{(0,0,v) \da} 
    \nonumber \\ 
    && \quad\quad\quad\quad  \to  e^{i (p_7 + p_8 + p_9) \pi} u^B_a b_{(0,0,u) \da} b_{(0,0,v) \ua}. 
\end{eqnarray}
This equation requires $u^B_a = e^{i (p_7 + p_8 + p_9)} u^B_a$. 

Similarly we define ${u^A_s{}^\ua = u^A_{(0,0,u), (0,0,v), \ua\ua}}$ 
and $u^A_s{}^\da = u^A_{(0,0,u), (0,0,v), \da\da}$. 
Acting $\sigma$ and $T_1^{-1}C_6^3$ on the $S$ term,
\begin{eqnarray}
&&    \sigma: u^A_s{}^\ua b^\dg_{(0,0,u)\ua} b^\pg_{(0,0,v)\ua} \nn
&& \quad\quad\quad  \quad\quad\quad    \to e^{i p_9 \pi} u^A_s{}^\ua b^\dg_{(0,0,v)\da} b^\pg_{(0,0,u)\da}, \\
&&    T_1^{-1}C_6^3: u^A_s{}^\ua b^\dg_{(0,0,u)\ua} b^\pg_{(0,0,v)\ua}  \nn
&& \quad\quad\quad   \quad\quad\quad   \to u^A_s{}^\ua b^\dg_{(0,0,v)\ua} b^\pg_{(0,0,u)\ua}.
    \label{eqn45}
\end{eqnarray}
From \eqnref{eqn45} we immediately conclude that if we require ${u^A_s{}^\ua \neq 0}$, 
then ${u^A_s{}^\ua = u^A_s{}^\ua{}^\ast}$, 
${u^A_s{}^\da = u^A_s{}^\da{}^\ast}$, 
and ${u^A_s{}^\ua = u^A_s{}^\da e^{i p_9 \pi}}$.

Applying $\sigma$ and $T_1^{-1}C_6^3$ on the ${u^B_s{}^\ua = u^B_{(0,0,u), (0,0,v), \ua\ua}}$ term, 
we see that
\begin{eqnarray}
  &&  \sigma: u^B_s{}^\ua b^\pg_{(0,0,u)\ua} b^\pg_{(0,0,v)\ua} \nn 
  &&  \quad\quad\quad\quad \to e^{i (p_7+1) \pi} u^B_s{}^\ua b^\pg_{(0,0,u)\da} b^\pg_{(0,0,v)\da}, \\
  && T_1^{-1}C_6^3: u^B_s{}^\ua b^\pg_{(0,0,u)\ua} b^\pg_{(0,0,v)\ua} \nn 
  &&  \quad\quad\quad\quad \to  e^{ i (p_7+p_8+p_9+1) \pi} u^B_s{}^\ua b^\pg_{(0,0,u)\ua} b^\pg_{(0,0,v)\ua}. 
\end{eqnarray}
and similarly for $u^B_s{}^\da$. It is obvious that such 
terms are nonzero only when ${p_7+p_8+p_9=1}$.

Following the same procedures, we find for $u^A_a$ terms
\begin{eqnarray}
    u^A_{(0,0,u), (0,0,v), \ua\da}{}^\ast &=& u^A_{(0,0,u), (0,0,v), \ua\da} e^{i p_9 \pi},\\
    u^A_{(0,0,u), (0,0,v), \ua\da} &=& -u^A_{(0,0,v), (0,0,u), \ua\da},
\end{eqnarray}
and similarly for $u^A_{(0,0,u), (0,0,v), \da\ua}$ and $u^A_{(0,0,v), (0,0,u), \da\ua}$.

We construct exchange interactions on all lattice bonds by applying symmetry operations. 
The results are shown in Tab.~\ref{stab2}.

\section{Fractionalization of crystal momentum and enhanced periodicity}
\label{appD}

Defining ${\mc{O}^s = \mc{G}^\mc{O} \mc{U}_\mc{O} \mc{O}}$ to be 
the symmetry group element acting on the spinon sector, we know 
from previous discussions that
\begin{eqnarray}
    T_1 \hat{T}_2 \hat{T}_1^{-1} \hat{T}_2^{-1} = (-1)^{p_1}.
\end{eqnarray}
Given a two-spinon product state ${\ket{a} = \ket{{\bs q}_a, \Omega_a}}$ 
with total momentum ${\bs q}_a$ and total energy $\Omega_a$, the 
translation operator acts on it by
\begin{eqnarray}
    \hat{T}_\mu\ket{a} = \hat{T}_\mu(1) \hat{T}_\mu(2) \ket{a} = e^{iq_a^\mu}\ket{a}
\end{eqnarray}
where $q_a^\mu = {\bs q}\cdot {\bs a}_\mu$.
We can construct another three states by translating the second spinon
\begin{eqnarray}
    \ket{b} &=& \hat{T}_1(2) \ket{a}, \\
    \ket{c} &=& \hat{T}_2(2) \ket{a}, \\
    \ket{d} &=& \hat{T}_1(2) \hat{T}_2(2) \ket{a}.  
\end{eqnarray}
These states have the same energy as $\ket{a}$, but with translated momenta,
\begin{eqnarray}
    (q_b^1, q_b^2) &=& (q_a^1, q_a^2+p_1 \pi),\\
    (q_c^1, q_c^2) &=& (q_a^1+p_1 \pi, q_a^2),\\
    (q_d^1, q_d^2) &=& (q_a^1+p_1 \pi, q_a^2+p_1 \pi).
\end{eqnarray}
Therefore, the two-spinon spectrum has an enhanced periodicity if $p_1 = 1$. 
In particular, the lower edge of $\mathcal{S}({\bs q},\omega)$
\begin{eqnarray}
    {\rm edge}(\bs q) = \min_{\rm k} \lz \omega({\bs k}) + \omega({\bs q- \bs k}) \rz
\end{eqnarray}
is completely encoded in energies of the two-spinons states with the momentum 
${\bs q}$, thus has the same periodicity,
\begin{eqnarray}
    {\rm edge}({\bs  q}_a) &=& {\rm edge}({\bs q}_b) 
    \nonumber \\
    &=& {\rm edge}({\bs  q}_c) = {\rm edge}({\bs q}_d). \label{piPeriod}
\end{eqnarray}
Otherwise ${p_1 = 0}$, and the lower excitation edge should have 
the usual periodicity of $2\pi$ in both directions of Brillouin zone basis.

We have shown that commuting and anticommuting single spinon translations 
gives different spectroscopic features. We now consider presentations of 
the symmetry group involving translations 
(since we are ultimately interested in the periodicity in the reciprocal space) 
on one-spinon sector,
\begin{eqnarray}
    \hat{T}_1^{-1} \hat{C}_6 \hat{T}_1 \hat{T}_2^{-1} \hat{C}_6^{-1} &=& (-1)^{p_2}, \label{eqn24}\\
    \hat{T}_2^{-1} \hat{C}_6 \hat{T}_1 \hat{C}_6^{-1} &=& (-1)^{p_3}, \label{eqn25}\\
    \hat{T}_1^{-1} \hat{\sigma} \hat{T}_1 \hat{\sigma}^{-1} &=& (-1)^{p_4}, \label{eqn26}\\
    \hat{T}_2^{-1} \hat{\sigma} \hat{T}_1 \hat{T}_2^{-1} \hat{\sigma}^{-1} &=& (-1)^{p_5}. \label{eqn27}
\end{eqnarray}
where we followed the convention in Appendix~\ref{appA}. Due to the gauge freedom, 
we can fix $p_2$ and $p_3$ to be $0$, and consistency of the PSG solution requires 
${p_4 = p_5 = p_1}$. With a detailed analysis, we found that \eqnref{eqn24}, 
\eqnref{eqn25} and \eqnref{eqn27} do not give to any obvious type of periodicity, 
while \eqnref{eqn26} gives a fuzzier version of the one constructed by considering 
translations only. Therefore, considering the whole symmetry group does not introduce 
more detailed implications of the neutron scattering spectrum.

\begin{figure}[t]
\centering
{
\includegraphics[height=4cm]{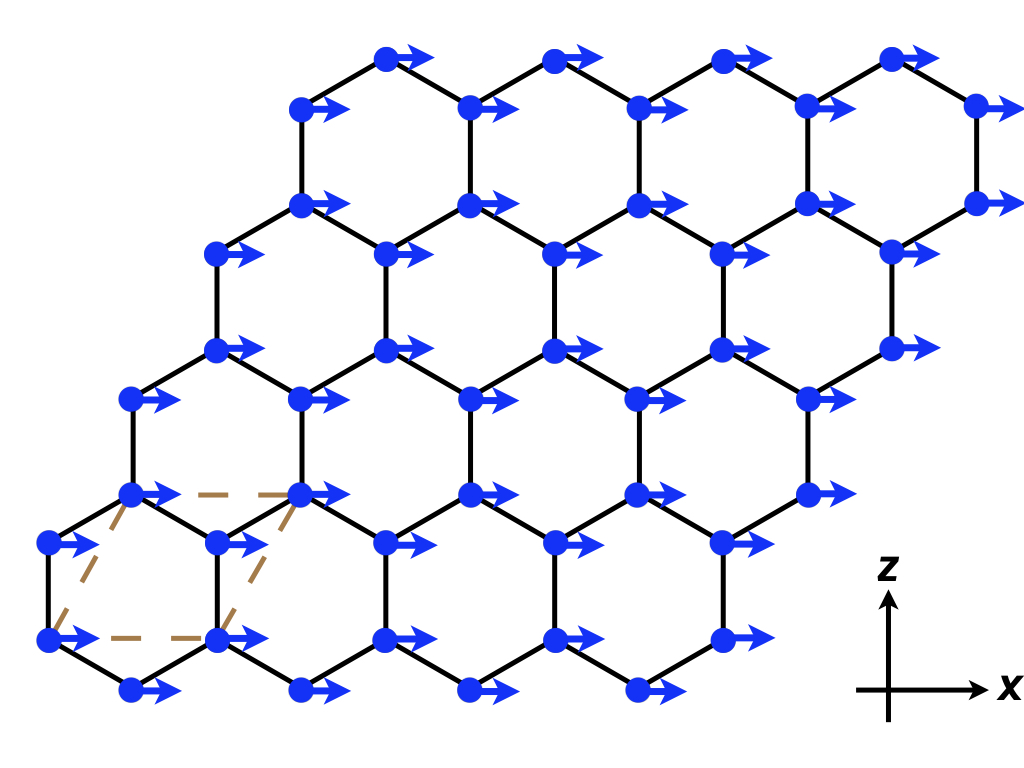}
}
\caption{
The typical magnetic ordering structure for a $\mathbb{Z}$2A100 state. 
The order is ferromagnetic, but the direction depends on the details 
of the condensate. The gray dashed lines denote the unit cell. 
The parameters of the Hamiltonian are the same as in Fig.~\ref{fig3}.
}
\label{sfig1}
\end{figure}

\begin{figure*}[t]
\centering
{
\includegraphics[width=4.5cm]{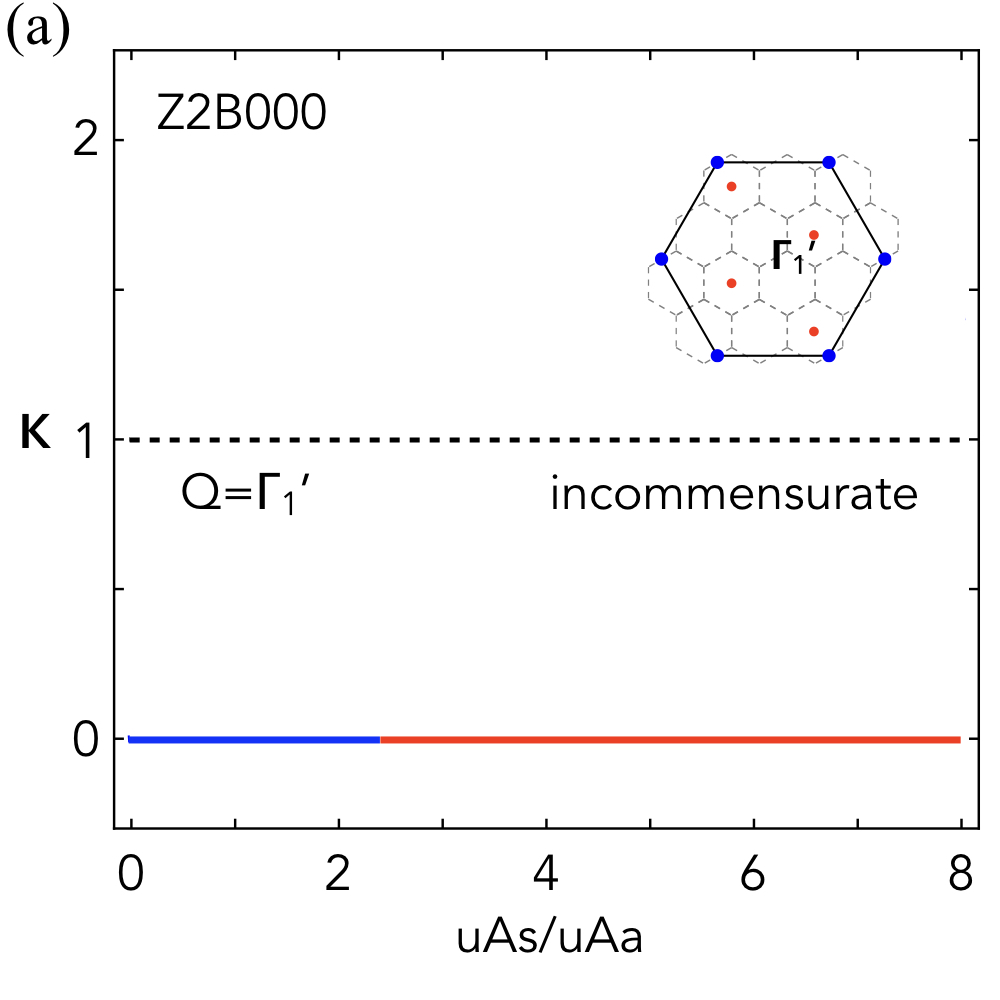}
\includegraphics[width=4.5cm]{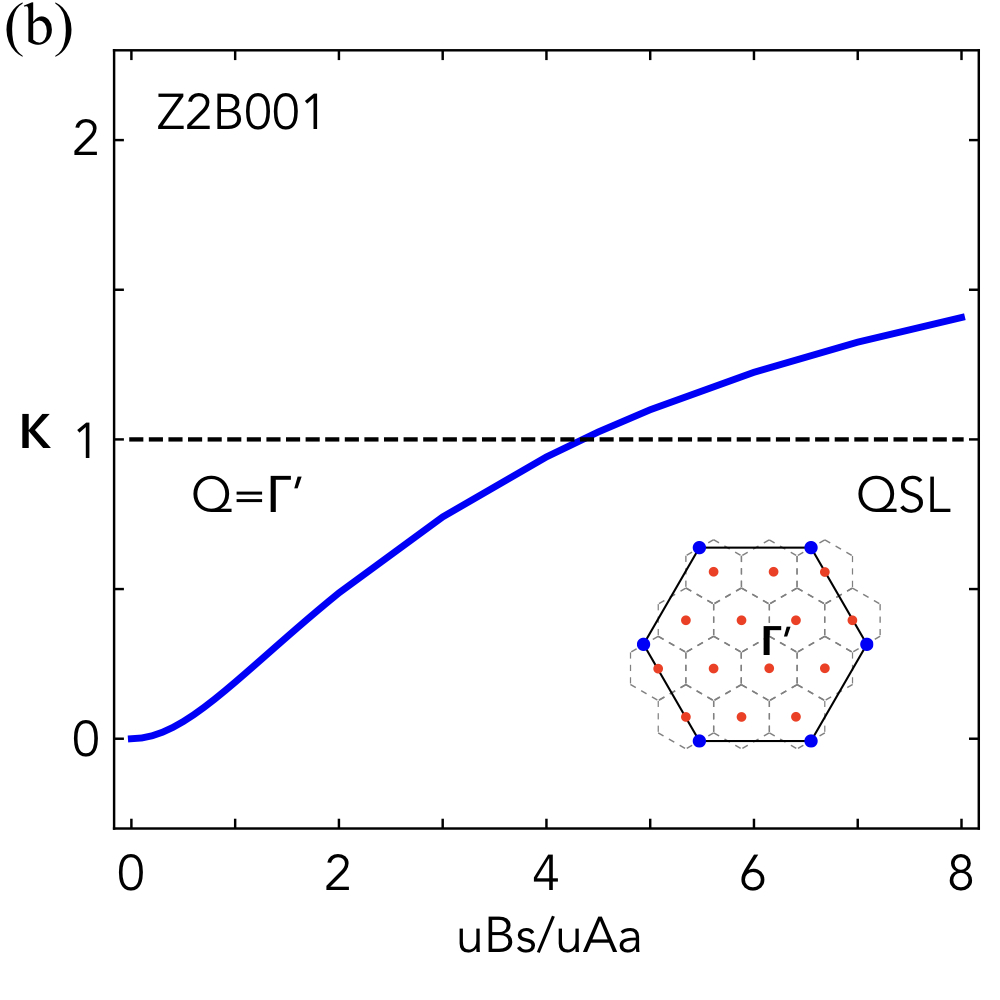}
\includegraphics[width=4.5cm]{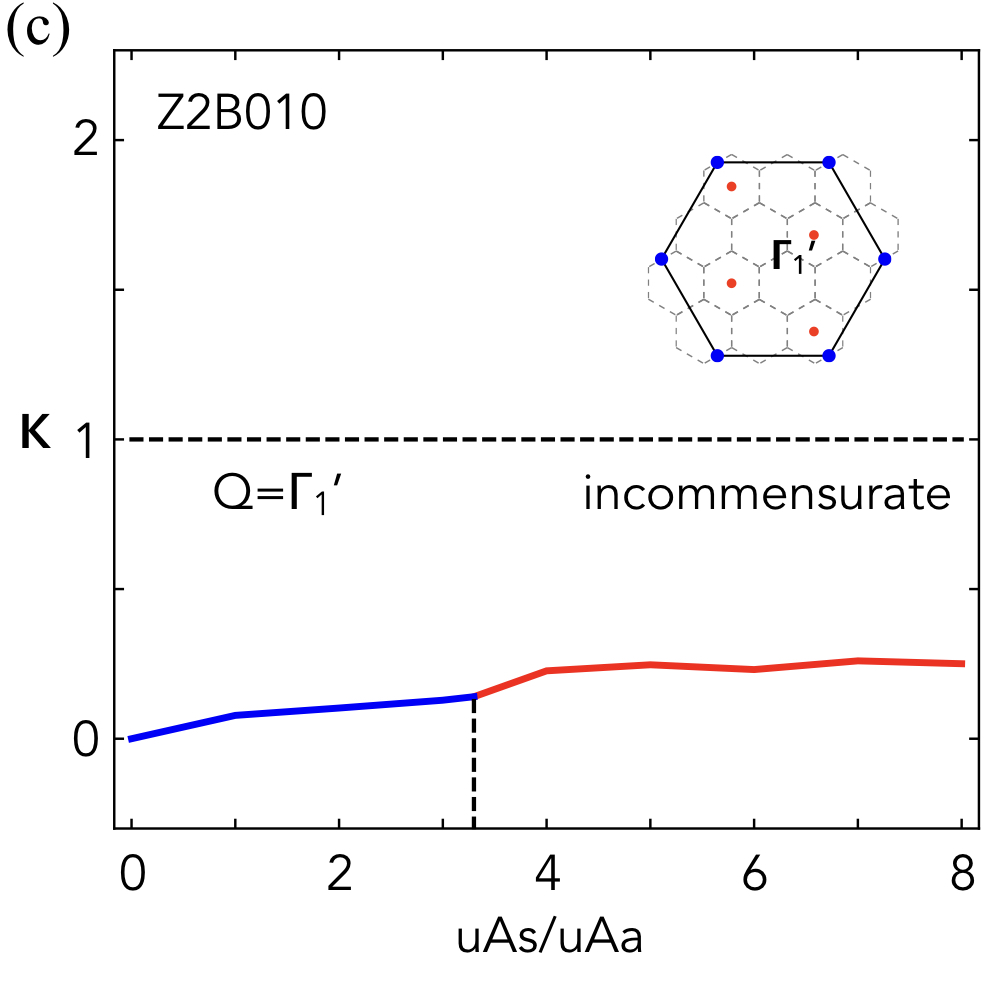}
\includegraphics[width=4.5cm]{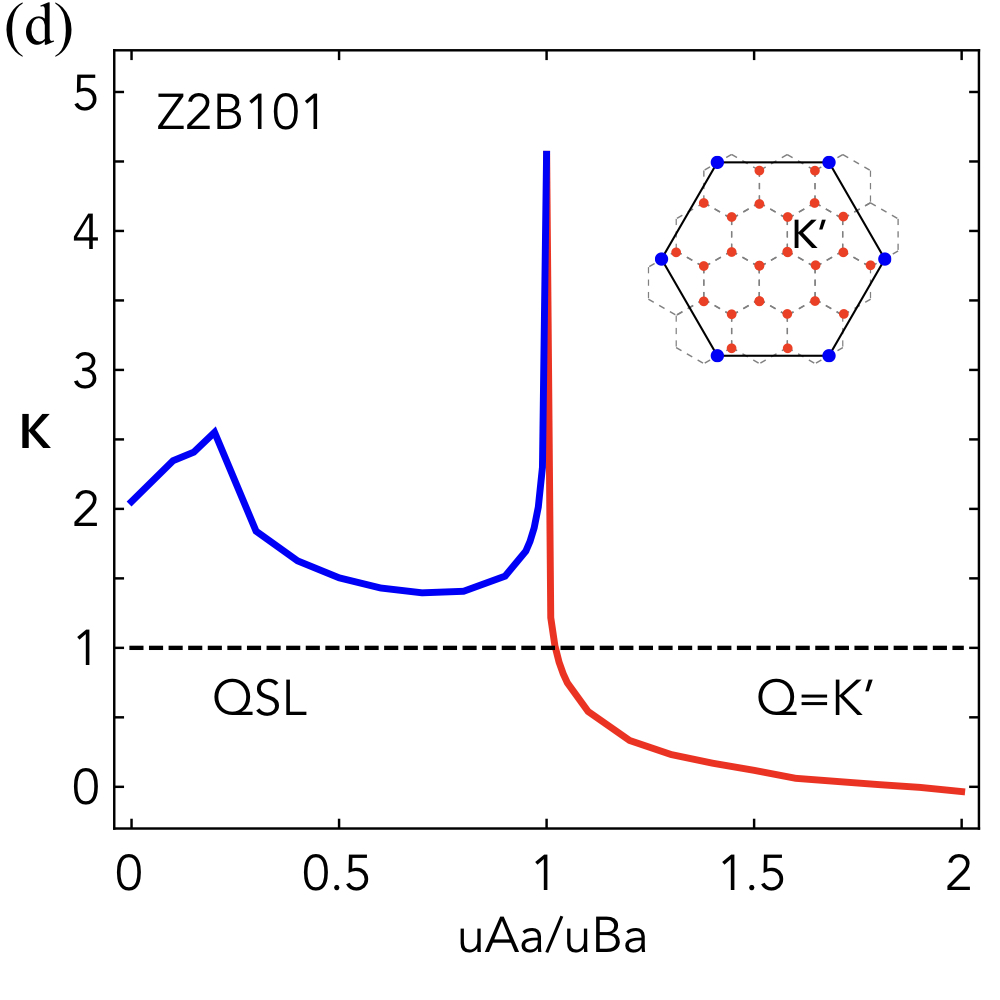}
\includegraphics[width=4.5cm]{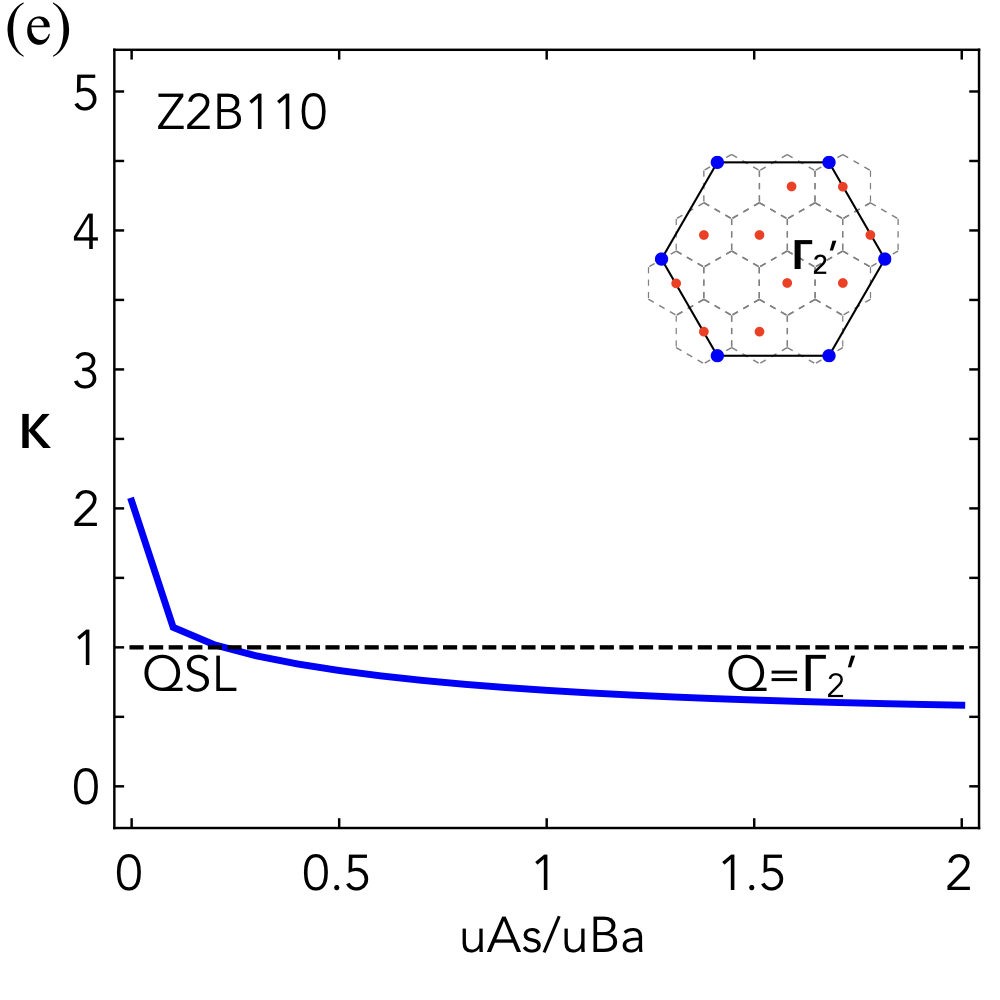}
}
\caption{(Color online.)
The phase diagrams for all other mean-field Hamiltonians in the $\mathbb{Z}$2B class. 
The nonzero parameters are listed in Tab.~\ref{tab1}. 
We choose $u^B_s/u^A_s=0.6$ in (c), and $u^A_a/u^A_s=0.6$ in (e).
}
\label{sfig2}
\end{figure*}

Rewriting Eq.~\eqref{eqn24}--\eqref{eqn27} in a more convenient form, 
and taking ${p_2 = p_3 = 0}$, ${p_4=p_5=p_1}$, we get
\begin{eqnarray}
    \hat{T}_1 \hat{C}_6 &=& \hat{C}_6 \hat{T}_1 \hat{T}_2^{-1}, \\
    \hat{T}_2 \hat{C}_6 &=& \hat{C}_6 \hat{T}_1, \\
    \hat{T}_1 \hat{\sigma} &=& (-1)^{p_1} \hat{\sigma} \hat{T}_1, \\
    \hat{T}_2 \hat{\sigma} &=& (-1)^{p_1} \hat{\sigma} \hat{T}_1 \hat{T}_2^{-1}.
\end{eqnarray}
Suppose ${\ket{a} = \ket{{\bs q}_a, \Omega_a}}$ is a two-spinon product state, 
we try acting $\hat{C}_6$ on the second spinon to obtain new eigenstates 
${\ket{b} = \hat{C}_6(2)\ket{a}}$ with the same energy. Then
\begin{eqnarray}
    T_1 \ket{b} &=& \hat{T}_1(1) \hat{T}_1(2) \hat{C}_6(2) \ket{a} \nn
                &=& \hat{C}_6(2) \hat{T}_1(1) \hat{T}_1(2) \hat{T}_2^{-1}(2) \ket{a} \nn
                &=& e^{i (q_a^1-k^2(2))} \ket{b}, \\
    T_2 \ket{b} &=& T_2^s(1) T_2^s(2) \hat{C}_6(2) \ket{a} \nn
                &=& \hat{C}_6(2) T_2^s(1) \hat{T}_1(2) \ket{a} \nn
                &=& e^{i (k^2(1)+k^1(2))} \ket{b},
\end{eqnarray}
where ${\bs k}(i)$ are the momenta for individual spinons. 
The result depends on the single spinon momentum, 
and does not lead to any obvious extra periodicity.

Similarly, let ${\ket{c} = \hat{\sigma}(2)\ket{a}}$,
\begin{eqnarray}
    T_1 \ket{c} &=& \hat{T}_1(1) \hat{T}_1(2) \hat{\sigma}(2) \ket{a} \nn
                &=& (-1)^{p_1} \hat{\sigma}(2) \hat{T}_1(1) \hat{T}_1(2) \ket{a} \nn
                &=& (-1)^{p_1} e^{i q_a^1} \ket{c}, \\
    T_2 \ket{c} &=& T_2^s(1) T_2^s(2) \hat{\sigma}(2) \ket{a} \nn
                &=& (-1)^{p_1} \hat{\sigma}(2) T_2^s(1) \hat{T}_1(2) \hat{T}_2^{-1}(2) \ket{a} \nn
                &=& (-1)^{p_1} e^{i (k^2(1)+k^1(2)-k^2(2))} \ket{c}.
\end{eqnarray}
While the second equation does not tell us much, the first one do carries 
$(q_a^1, q_a^2)$ to $(q_c^1, q_c^2)$ with ${q_c^1 = q_a^1 + p_1 \pi}$, 
while we cannot say much about $q_c^2$ and $q_a^2$. This is a fuzzier 
version of \eqnref{piPeriod} as it does not carry as much information 
about the structure of the spectrum.

\section{Proximate magnetic order of $\mathbb{Z}_2$ QSLs}
\label{appE}

\subsection{$\mathbb{Z}$2A states}

Here we briefly comment on the proximate magnetic order resulting from the $\mathbb{Z}$2A parent state. In this case, the translation symmetry is not fractionalized, and the proximate magnetic order also preserves such a symmetry. For a large range of parameters, the band minimum is at ${\bs{\Gamma} = (0,0)}$, giving rise to a ferromagnetic order.
The ordering pattern for a typical set of parameters is shown in Fig.~\ref{sfig1}.

\subsection{$\mathbb{Z}$2B states}
\label{appEB}

In this section we discuss other $\mathbb{Z}$2B mean field classes in the PSG classification. 
Among them the $\mathbb{Z}$2B011 class has only one nonzero parameter $u^A_a$, and the corresponding 
ground state is magnetically ordered with ordering wave vector $\bs{\Gamma}_1^\p$ 
and $\bs{\Gamma}_2^\p$. Phase diagrams of the other classes are shown in Fig.~\ref{sfig2}.

We see that for the parameter space we choose, the $\mathbb{Z}$2B000 and $\mathbb{Z}$2B010 states are always ordered, 
while $\mathbb{Z}$2B001, $\mathbb{Z}$2B101, and $\mathbb{Z}$2B110 can all support a QSL phase. The ordering wave 
vector $\bs{\Gamma}_1^\p$ for $\mathbb{Z}$2B000 and $\mathbb{Z}$2B010 are the same as that of thef $\mathbb{Z}$2B100 described 
and discussed in Sec.~\ref{sec3} and Sec.~\ref{sec4}, and thus consistent with the magnetic 
Bragg peak at $(\pi, 0)$.

\bibliography{refs}

\end{document}